\newcommand{\rem}[1]{}
\documentclass[prb,twocolumn,showpacs,preprintnumbers,amsmath,amssymb,floatfix]{revtex4}
\usepackage{hyperref}
\usepackage{epsfig}

\begin{document}

\title{Adiabatic quantum dynamics of a random Ising chain across its quantum critical point}

\author{Tommaso Caneva}
\affiliation{International School for Advanced Studies (SISSA), Via Beirut 2-4,
  I-34014 Trieste, Italy}

\author{Rosario Fazio}
\affiliation{International School for Advanced Studies (SISSA), Via Beirut 2-4,
  I-34014 Trieste, Italy}
\affiliation{NEST-CNR-INFM $\&$ Scuola Normale Superiore di Pisa, 
  Piazza dei Cavalieri 7, 56126 Pisa, Italy}

\author{Giuseppe E. Santoro}
\affiliation{International School for Advanced Studies (SISSA), 
  Via Beirut 2-4, I-34014 Trieste, Italy}
\affiliation{CNR-INFM Democritos National Simulation Center, 
  Via Beirut 2-4, I-34014 Trieste, Italy}
\affiliation{International Centre for Theoretical Physics (ICTP), 
  P.O.Box 586, I-34014 Trieste, Italy}

\date{\today}

\begin{abstract}
We present here our study of the adiabatic quantum dynamics of a random Ising chain across its
quantum critical point.
The model investigated is an Ising chain in a transverse field with disorder present both in the exchange 
coupling and in the transverse field. The transverse field term is proportional to a function $\Gamma(t)$
which, as in the Kibble-Zurek mechanism, is linearly reduced to zero in time with a rate $\tau^{-1}$, 
$\Gamma(t)=-t/\tau$, starting at $t=-\infty$ from the quantum disordered phase ($\Gamma=\infty$) and ending 
at $t=0$ in the classical ferromagnetic phase ($\Gamma=0$). 
% HERE ELIMINATED HYPHENS
We first analyze the distribution of the gaps, occurring at the critical point $\Gamma_c=1$, which 
are relevant for breaking the adiabaticity of the dynamics. 
We then present extensive numerical simulations for the residual energy $E_{\rm res}$ and density of defects
$\rho_k$ at the end of the annealing, as a function of the annealing inverse rate $\tau$. 
%for different lenghts of the chain.
Both the average $E_{\rm res}(\tau)$ and $\rho_k(\tau)$ are found to behave logarithmically for large $\tau$, 
but with different exponents, $[E_{\rm res}(\tau)/L]_{\rm av}\sim 1/\ln^{\zeta}(\tau)$ with $\zeta\approx 3.4$, and
$[\rho_k(\tau)]_{\rm av}\sim 1/\ln^{2}(\tau)$. 
We propose a mechanism for $1/\ln^2{\tau}$-behavior of $[\rho_k]_{\rm av}$ based on the Landau-Zener tunneling theory 
and on a Fisher's type real-space renormalization group analysis of the relevant gaps. 
The model proposed shows therefore a paradigmatic example of how an adiabatic quantum computation can become
very slow when disorder is at play, even in absence of any source of frustration.
%We finally show that the introduction of an anisotropy parameter in the plane 
%orthogonal to the field leaves unaltered the character of the transition.        
\end{abstract}

%\pacs{
%}

\maketitle

%++++++++++++++++++++++++++++++++++++++++++++++++++++++++++++++++++++++++++++++++++
\section{Introduction}
%++++++++++++++++++++++++++++++++++++++++++++++++++++++++++++++++++++++++++++++++++
%
How effective is it to execute a given computational task by slowly varying in time 
the Hamiltonian of a quantum system? Is it possible to find the ground state of a classical 
system by slowly annealing away its quantum fluctuations? What is the density of defects 
left over after a passage through a continuous (quantum) phase transition? These seemingly 
different problems, rubricated under the names of Adiabatic quantum computation~\cite{Farhi_SCI01}, Quantum 
% HERE ADDED REF.
Annealing~\cite{Finnila_CPL94,Kadowaki_PRE98,Brooke_SCI99,Santoro_SCI02,Das_Chakrabarti:book,Santoro_JPA:review} 
and Kibble-Zurek topological defect formation in quantum phase transitions~\cite{Zurek_PRL05}, 
are ultimately related to the understanding of the failure of the adiabatic approximation in a 
many-body system passing through a quantum critical point. 

Adiabatic quantum computation (AQC), {\it alias} Quantum Annealing (QA), is a possible alternative to 
the standard circuit-theory approach to Quantum Computation (QC)~\cite{Nielsen_Chuang:book}. 
Indeed, as shown by Aharonov {\it et al.}~\cite{Aharonov:proceeding,Aharonov_QP04:preprint}, 
any quantum algorithm can be equivalently reformulated in terms of the adiabatic evolution of an appropriate 
time-dependent Hamiltonian $H(t)=[1-f(t)]H_{\rm in}+f(t)H_{\rm fin}$, 
$f(t)$ being a generic function of time such that $f(0)=0$ and $f(t_{\rm fin})=1$.
The initial Hamiltonian $H_{\rm in}$, for which we know the ground state, provides the 
input of the algorithm. The final Hamiltonian $H_{\rm fin}$ is constructed appropriately so as
to possess the solution of the computational task as its ground state. 
The knowledge of the equivalence of computational power between the two
different QC schemes, however, does not provide a practical way of constructing
$H_{\rm in}$ and $H_{\rm fin}$ for a given computational problem.
Understanding what computational problems can be efficiently solved by AQC-QA is, 
in general, a very difficult problem. In order to solve the task one has to find a 
suitable path in Hamiltonian space in such a way that the resulting Schr\"odinger 
evolution efficiently drives the system from some simple initial quantum state 
$|\Psi_{\rm in}\rangle$ to the sought final ground state~\cite{footnote_gamma,Roland_PRA02}.
The accuracy of the computation, which relies on the possibility for the system to 
remain in the instantaneous ground state during the dynamics, is ultimately limited by the fact that 
at specific times the instantaneous Hamiltonian presents a gap between the ground and the first excited 
state which closes on increasing the size of the input. 

On totally independent grounds, the study of topological defect formation 
goes back to the 80's, motivated by the effort to understand signatures of phase 
transitions which have occurred in the early universe~\cite{Kibble:review,Zurek:review} by 
determining the density of defects left in the broken symmetry phase as a function of 
the rate of quench.
By means of the so called Kibble-Zurek mechanism, a scaling law relates the density 
of defects to the rate of quench. 
The suggestion of Zurek to simulate transitions in the early universe by means of condensed 
matter system has stimulated an intense experimental activity~\cite{Bauerle_NAT96,Ruutu_NAT96}
aimed at verifying the Kibble-Zurek theory. 
The understanding of defect formation was later explored also in the case of a quantum phase 
transition~\cite{Zurek_PRL05,Polkovnikov_PRB05}, where the crossing of the critical point is done by 
varying a parameter in the Hamiltonian. 
These works have stimulated an intense activity where several different quantum systems undergoing a quantum phase
transition were scrutinized.
In the past couple of years there have been a number of results obtained in the area of adiabatic 
dynamics of many-body systems~\cite{footnote_Polkovnikov,Polkovnikov_07:preprint}. 
Most of the works concentrated on the one-dimensional Ising model. 
Soon after the appearance of Ref.~\onlinecite{Zurek_PRL05}, Dziarmaga~\cite{Dziarmaga_PRL05}
obtained analytically the scaling law for the density of defects by resorting to 
the exact solution of Pfeuty~\cite{Pfeuty_AP70}. 
A detailed analysis {\it a' la Landau-Zener} was presented in Refs.~\cite{Damski_PRL05,Damski_PRA06,Cherng_PRA06}.
The effect of an external noise on the adiabatic evolution and its consequences for 
the Kibble-Zurek mechanism has been discussed in~\cite{Fubini_07:preprint}. 
Recently, quenches in Bose-Hubbard models were analyzed~\cite{Schutzhold_PRL06,Cucchietti_PRA07} as well.
Observables which were analyzed to quantify the loss of adiabaticity in the critical region were typically 
%the residual energy, 
the density of defects left behind in the broken symmetry phase, the fidelity of the evolved state with 
respect to the ground state, and, in few cases, also the residual block entropy~\cite{Latorre_PRA04,Cincio_PRA07}. 
This brief overview of recent works accounts only for papers dealing with adiabatic dynamics, without
touching the vast literature treating the case of sudden quenches. 

In the present work we analyze the adiabatic dynamics in a one-dimensional quantum 
disordered Ising model in a random transverse field. 
The reasons for considering this problem are various. 
First of all it is an important ground test for the Kibble-Zurek mechanism. 
In addition, although in a very simplified manner, it may help 
in understanding more interesting problems that can be formulated in terms of interacting Ising spins, 
Traveling Salesman~\cite{Hopfield_SCI86} and Satisfiability~\cite{Mezard_SCI02} problems being only 
two well-known examples. 
The simplicity of our test problem lies in the particularly simple geometry of the 
interactions, which forbids frustration. The only ingredient that our problem shares 
with more challenging computational tasks is the fact that the interactions are chosen to be random. 
This feature, the presence of disorder, makes the problem interesting and non-trivial for a physically 
inspired computational approach based on AQC-QA. 

Of particular relevance for us is Ref.\onlinecite{Dziarmaga_PRB06} where this model was 
analyzed first, and the anomalously slow dynamics characterized by an average 
density of kinks which vanishes only logarithmically with the annealing rate. 
Here we extend this work by presenting a detailed analysis of the statistics 
of both the residual energy and kink density. In a disordered chain, the formation 
of the kinks is no longer translational invariant and therefore it affects in a 
non-trivial way, as we will show below, the scaling of the residual energy. 

The rest of the paper is organized as follows: 
In Sec.~\ref{model:sec} we define the problem and the technique to solve the adiabatic dynamics of
the random Ising chain, and next, in Sec.~\ref{residual:sec}, we introduce the quantities --- residual
energy and density of defects --- that we calculate to quantify the departure from the adiabatic ground state. 
In Sec.~\ref{results:sec} we present our numerical results for both these quantities, together with an 
analysis of the large-annealing-time behavior of the density of defects, based on the Landau-Zener theory, 
explicitly showing the slow dynamics which the disorder entails.
In the final section we present a critical assessment of our findings, and a concluding discussion.

%++++++++++++++++++++++++++++++++++++++++++++++++++++++++++++++++++++++++++++++++++ 
\section{The model} \label{model:sec}
%++++++++++++++++++++++++++++++++++++++++++++++++++++++++++++++++++++++++++++++++++
%
As discussed in the Introduction, our aim is to analyze the adiabatic dynamics of 
a one-dimensional random Ising model defined by the Hamiltonian
\begin{equation} \label{random_ising:eqn}
H(t) = -\sum _i J_i \sigma^z_i \sigma^z_{i+1} - \Gamma(t) \sum_i h_i \sigma^x_i\;.
\end{equation}
In the previous expression $\sigma^{\alpha}_i$ ($\alpha = x,z$) are Pauli matrices 
for the $i$-th spin of the chain, $J_i$ are random couplings between neighboring spins, and
$h_i$ are random transverse fields.
The time-dependent function $\Gamma(t)$ rescaling the transverse field term allows
us to drive the system form a region of infinitely high transverse fields ($\Gamma=\infty$, where
the ground state has all spins aligned along $x$, see below), to the case of a classical
Ising model ($\Gamma=0$). 
Specifically, we will take in the following $\Gamma(t)$ to be a linear function of time characterized by an
annealing rate $\tau^{-1}$
\[
\Gamma(t) = -\frac{t}{\tau} \hspace{6mm} \mbox{for} \; t \in (-\infty,0] \;\;.
\]
In one-dimension, and for nearest-neighbor couplings, there is 
no frustration associated to the random nature of the couplings $J_i$: by appropriately 
performing spin rotations of $\pi$ along the $x$-spin axis, we can always change the 
desired $\sigma^z_i$ into $-\sigma^z_i$ and invert accordingly the signs of the couplings 
in such a way that all $J_i$'s turn out to be non-negative. We therefore assume that 
the $J_i$ are randomly distributed in the interval $[0,1]$, specifically with a flat 
distribution $\pi[J]=\theta(J)\theta(1-J)$, where $\theta$ is the Heaviside function.
The same distribution is used for the random field $\pi[h]=\theta(h)\theta(1-h)$. 
This is different from the model considered in Ref.\onlinecite{Dziarmaga_PRB06}, where the disorder 
was introduced in the exchange coupling only. We find the present choice quite convenient since,
by duality arguments~\cite{Fisher_PRB95}, the critical point separating the large-$\Gamma$ quantum 
paramagnetic phase from the low-$\Gamma$ ferromagnetic region is known to be located at $\Gamma_c=1$.

At the initial time $t_{\rm in}=-\infty$ the ground state of $H(t_{\rm in})$, completely dominated
by the transverse field term, is simply the state with all spins aligned along the $+\hat{x}$ spin direction: 
$|\Psi_{\rm in}\rangle=\prod_i |\hat{x}\rangle_i 
= \prod_i [|\!\!\uparrow\rangle_i+|\!\!\downarrow\rangle_i]/\sqrt{2}$. 
On the other side of the transition point $\Gamma_c$, the final Hamiltonian $H(t_{\rm fin})=H_{cl}$
describes a {\em random ferromagnet} whose ground states,
% HERE ELIMINATED HYPHEN
which we aim to reach by adiabatically switching off $\Gamma(t)$, are obviously 
the two trivial states $|\Psi_{\uparrow}\rangle=\prod_i|\!\! \uparrow\rangle_i$ and 
$|\Psi_{\downarrow}\rangle=\prod_i|\!\! \downarrow\rangle_i$: 
as an optimization problem, $H_{\rm fin}$ represents, therefore, a trivial problem. 

Even if the ground states in the two limiting cases, $\Gamma=\infty$ and $\Gamma=0$, are very easy to find, 
when it comes to dynamics, the evolution dictated by $H(t)$ is no longer a trivial problem. 
The instantaneous spectrum of the Hamiltonian $H(t)$ 
% HERE CHANGED
% possesses a whole region, around and below $\Gamma_c$, which 
is gapless in the thermodynamic limit \cite{Fisher_PRB95}. 
This implies that, during the adiabatic evolution, defects in the form of domain walls between differently aligned
ferromagnetic ground states, of the type 
\begin{equation} \label{domains:eqn}
|\dots\uparrow\downarrow\downarrow\downarrow\downarrow\downarrow
\uparrow\uparrow\uparrow\uparrow\uparrow\uparrow\uparrow\uparrow
\downarrow\downarrow\downarrow\downarrow\dots\rangle \nonumber
\end{equation}
are formed, and reflected in a whole structure of closing gaps will appear in the instantaneous spectrum.

%%++++++++++++++++++++++++++++++++++++++++++++++++++++++++++++++++++++++++++++++++++
\subsection{Fermion representation and Bogoliubov-de Gennes equations}
\label{method:sec}
%++++++++++++++++++++++++++++++++++++++++++++++++++++++++++++++++++++++++++++++++++
%
By means of the Jordan-Wigner transformation, the one-dimensional Ising model is reduced 
to a free fermion model. One first writes the spin operators in terms of hard-core bosons $a_i$ and 
$a_i^{\dagger}$ in a representation that maps the state $|\sigma^z_i=+1\rangle \to 
|1\rangle_i=a^{\dagger}_i|0\rangle_i$  and $|\sigma^z_i=-1\rangle \to |0\rangle_i$, 
with the hard-core constraint $(a^{\dagger}_i)^2|0\rangle_i=0$:
$\sigma_i^z = 2a^{\dagger}_i a_i-1$, $\sigma_i^x = a_i+a^{\dagger}_i$, and 
$\sigma_i^y = -i(a^{\dagger}_i-a_i)$.
The hard-core boson operators $a_i$ are then re-expressed in terms of spinless fermions 
operators $c_i$: $a_i = e^{i\pi\sum_{j<i} c^{\dagger}_j c_j} c_i$.
After a $\pi/2$ rotation around the y-axis, which maps $\sigma^x\to \sigma^z$ and $\sigma^z\to -\sigma^x$,
the Hamiltonian in Eq.(\ref{random_ising:eqn}) can be rewritten in terms of fermion operators as 
\begin{equation} \label{H_OBC}
H = - \sum_i^{L-1} J_i \{c^{\dagger}_i c^{\dagger}_{i+1} + c^{\dagger}_i c_{i+1}  + {\rm H.c.}\} 
    - 2\Gamma \sum_i^{L} h_i c^{\dagger}_i c_i \;,
\end{equation}
where we have assumed open boundary conditions (OBC) for the spin-chain. 
For the case of periodic boundary conditions (PBC) on the spins, $\sigma_{L+1}=\sigma_1$, 
extra boundary terms appear in the fermionic Hamiltonian, of the form 
$\Delta H_{\rm PBC} = J_L (-1)^{N_F} \{ c^{\dagger}_L c^{\dagger}_{1} 
                                      + c^{\dagger}_L c_{1} + {\rm H.c.} \}$, 
where $N_F=\sum_i c^{\dagger}_i c_i$ is the total number of fermions. 
Notice that although $N_F$ is not conserved by the Hamiltonian (\ref{H_OBC}), the parity 
of $N_F$ is conserved: $(-1)^{N_F}$ is a constant of motion with value $1$ or $-1$. 

%..........................................................
\subsubsection{Statics}
%..........................................................
%
The model in Eq.~(\ref{H_OBC}) can be diagonalized through a Bogoliubov 
rotation~\cite{Lieb_AP61,Young_PRB97}, 
by introducing the new fermionic operators $\gamma_{\mu}$ and $\gamma^{\dagger}_{\mu}$ 
\begin{eqnarray} \label{gamma_c_transf:eqn}
\gamma_{\mu} &=& \sum^L_{j=1} (u_{j\mu}^* c_j + v^*_{j\mu} c^{\dagger}_j)  \nonumber \\
%\gamma^{\dagger}_{\mu} &=& \sum^L_{j=1} (v_{j\mu} c_j + u_{j\mu} c^{\dagger}_j) \;,
c_{i} &=& \sum_{\mu=1}^{L} ( u_{i\mu}  \gamma_{\mu} + v_{i\mu}^* \gamma_{\mu}^{\dagger} ) \;,
\end{eqnarray}
where the L-dimensional vectors ${\bf u}_{\mu}$ and ${\bf v}_{\mu}$, for $\mu=1,\cdots,L$, satisfy the 
Bogoliubov-de Gennes equations:
\begin{eqnarray} \label{BdG_static:eqn}
 A \cdot {\bf u}_{\mu} + B \cdot {\bf v}_{\mu} &=& \epsilon_{\mu} {\bf u}_{\mu} \nonumber \\
-B \cdot {\bf u}_{\mu} - A \cdot {\bf v}_{\mu} &=& \epsilon_{\mu} {\bf v}_{\mu} \;.
\end{eqnarray}
Here $A$ and $B$ are real $L\times L$ matrices whose non-zero elements are given by 
$A_{i,i}=-\Gamma h_i$, $A_{i,i+1}=A_{i+1,i}=-J_i/2$, $B_{i,i+1}=-B_{i+1,i}=-J_i/2$.
(For the PBC spin-chain case, we have the additional matrix elements
$A_{L,1}=A_{1,L}=(J_L/2)(-1)^{N_F}$, and $B_{L,1}=-B_{1,L}=(J_L/2)(-1)^{N_F}$). 
While in the ordered case the solution of Eqs.(\ref{BdG_static:eqn}) can be reduced, by switching to momentum-space, 
to independent $2\times 2$ problems, in the general disordered case one has to diagonalize the $2L \times 2L$ problem
numerically \cite{Young_PRB96,Fisher_PRB98}. 

The spectrum of Eqs.~(\ref{BdG_static:eqn}) turns out to be given by $\pm \epsilon_{\mu}$, with 
$\epsilon_{\mu}\ge 0$, and in terms of the new fermion operators, $H$ becomes:
\begin{equation}
H=\sum_{\mu=1}^{L}(\epsilon_{\mu}\gamma^{\dagger}_{\mu}\gamma_{\mu}-\epsilon _{\mu}\gamma_{\mu}\gamma^{\dagger}_{\mu})
 =\sum_{\mu=1}^{L} 2\epsilon_{\mu}(\gamma_{\mu}^{\dagger}\gamma_{\mu}-\frac{1}{2}) \;.
\end{equation}
The ground state of $H$ is the Bogoliubov vacuum state $|\Psi_0\rangle$
annihilated by all $\gamma_{\mu}$ for $\mu=1\cdots L$, $\gamma_{\mu}|\Psi_0\rangle=0$, with an energy
$E_0=-\sum_{\mu=1}^{L} \epsilon_{\mu}$. 

%.................................................
\subsubsection{Dynamics} 
%.................................................
%
The Schr\"odinger dynamics associated to a time-dependent $H(t)$ can be solved 
by a time-dependent Bogoliubov theory~\cite{Barouch_PRA70}. 
The basic fact that makes the solution possible even in the time-dependent case is that the  
Heisenberg's equations of motion for the operators $c_{i,H}(t)$ are {\em linear}, because the
Hamiltonian is quadratic: 
\begin{equation} \label{eq_mot_c:eqn}
i\hbar \frac{d}{dt} c_{i,H}(t) 
= 2\sum_{j=1}^{L} \left[A_{i,j}(t) c_{j,H}(t) + B_{i,j}(t) c_{j,H}^{\dagger}(t)\right] \;.
\end{equation}
Here the matrices $A$ and $B$ have the same form given previously, except that now the time-dependence of $\Gamma(t)$ 
is explicitly accounted for. If we denote by $\gamma_{\mu,{\rm in}}$ the Bogoliubov operators that 
diagonalize $H(t_{\rm in})$ at the initial time, and ${\bf u}_{\mu}^{\rm in}$, ${\bf v}_{\mu}^{\rm in}$ 
the corresponding initial eigenvectors, it is simple to verify that the {\em Ansatz}
\begin{equation} \label{time_dep_Bog_trans:eqn}
c_{i,H}(t) = \sum_{\mu=1}^{L} \left( u_{i\mu}(t)   \gamma_{\mu,{\rm in}}
                                   + v_{i\mu}^*(t) \gamma_{\mu,{\rm in}}^{\dagger} \right) \;,
\end{equation} 
does indeed solve the Heisenberg equations (\ref{eq_mot_c:eqn}), provided the time-dependent
coefficients $u_{i\mu}(t)$ and $v_{i\mu}(t)$, satisfy the following system of first-order differential equations
\begin{eqnarray} \label{BdG_tdep:eqn}
i\frac{d}{dt}u_{i\mu}(t) \!\! &=&  
\frac{2}{\hbar} \sum_{j=1}^{L} \left[A_{i,j}(t)u_{j\mu}(t)+B_{i,j}(t)v_{j\mu}(t) \right] 
\nonumber \\
i\frac{d}{dt}v_{i\mu}(t) \!\! &=& \!\! 
-\frac{2}{\hbar} \sum_{j=1}^{L} \left[A_{i,j}(t)v_{j\mu}(t)+B_{i,j}(t)u_{j\mu}(t) \right] 
\;,
\end{eqnarray}
with initial condition $u_{i\mu}(t_{\rm in})=u_{i\mu}^{\rm in}$, $v_{i\mu}(t_{\rm in})=v_{i\mu}^{\rm in}$. 
Eqs.~(\ref{BdG_tdep:eqn}) are the natural time-dependent generalizations of the static Bogoliubov-de Gennes 
Eqs.~(\ref{BdG_static:eqn}), and, once again, they have to be solved numerically in the general disordered
case. 

%--------------------------------------------------------------------
\section{Residual energy and kink density} \label{residual:sec}
%--------------------------------------------------------------------
%
How effectively the Schr\"odinger dynamics drives the system from the initial disordered quantum ground state 
$|\Psi_{\rm in}\rangle$ towards the classical ground state 
$|\Psi_{\uparrow}\rangle=\prod_i|\!\! \uparrow\rangle_i$ 
(or the fully reversed one $|\Psi_{\downarrow}\rangle=\prod_i|\!\! \downarrow\rangle_i$)? 

A way of quantifying the degree of adiabaticity of the evolution is given by the residual energy, defined as 
\begin{equation}
E_{\rm res}=E_{\rm fin}-E_{\rm cl} \;,
\end{equation} 
where $E_{\rm cl}=-\sum_i J_i$ is the classical ground state energy of $H(t_{\rm fin})=H_{\rm cl}$, and 
$E_{\rm fin}=\langle \Psi_{\rm fin}|H_{\rm cl}|\Psi_{\rm fin}\rangle$ 
is the average classical energy of the final time-evolved state $|\Psi_{\rm fin}\rangle$.
% HERE ELIMINATED HYPEHNS
Obviously, $E_{\rm fin}$, and hence $E_{\rm res}$, depends on the parameters 
specifying the evolution: the smaller and closer to $E_{\rm cl}$ the ``slower'' the evolution.

An alternative way of quantifying the degree of adiabaticity of the evolution, is given in 
terms of the density of kinks $\rho_k$ in the final state, defined by
\begin{equation} \label{ddefects}
\rho_k=\frac{1}{L} \sum_{i}^{L-1} \langle \Psi(0)| \frac{1}{2} \left(1-\sigma_i^{z}\sigma_{i+1}^{z}\right)
|\Psi(0) \rangle 
\end{equation}
(for a PBC chain the sum goes up to $L$, instead of $L-1$). 

When no disorder is present the two quantities coincide, apart from trivial constants. 
In the disordered case, however, this is not the case. A defect will form with higher probability 
at a link where the corresponding exchange coupling $J_i$ is small. Therefore the residual energy is 
not simply given by the kink density times the exchange coupling.

The calculation of quantities like $E_{\rm fin}$ or $\rho_k$ is straightforward.
Quite generally, given an operator $\hat{O}[c_i,c^{\dagger}_i]$ expressed in terms 
of the $c_i$'s and $c^\dagger_i$'s,
its expectation value over the final state $|\Psi(t_{\rm fin}=0)\rangle$ can be expressed, 
switching from the Schr\"odinger to the Heisenberg picture, as
$\langle \Psi(0)| \hat{O}[c_i,c^\dagger_i] |\Psi(0)\rangle = 
 \langle \Psi(t_{\rm in})| \hat{O}[c_{i,H}(0),c^\dagger_{i,H}(0)] |\Psi(t_{\rm in})\rangle$.
Next, one uses the expressions (\ref{time_dep_Bog_trans:eqn}) for the $c_{i,H}(0)$'s 
and $c_{i,H}^{\dagger}(0)$ in terms of $\gamma_{\mu,in}$, $\gamma^{\dagger}_{\mu,in}$, 
$u_{i,\mu}(0)$, and $v_{i,\mu}(0)$,  
and uses the fact that the $\gamma_{\mu,in}$ annihilates by construction the initial state 
$|\Psi(t_{\rm in})\rangle$.

By applying this procedure to the calculation of $E_{\rm fin}$ we get:
\begin{eqnarray}
E_{\rm fin} &=& \sum_{i,j} 
\left( A_{ij}(0) \left[ v(0)v^{\dagger}(0) - u(0)u^{\dagger}(0) \right]_{ij} 
+\right. \nonumber \\
&&\hspace{7mm} \left. B_{ij}(0) \left[ v(0)u^{\dagger}(0) - u(0)v^{\dagger}(0) \right]_{ij}
\right) \;, 
\end{eqnarray}
where $u(0)$ and $v(0)$ are $L\times L$ matrices with elements $u_{i,\mu}(0)$ and $v_{i,\mu}(0)$.
Similarly, the density of defects $\rho_k$ can also be expressed as:
\begin{equation}
\rho_k = \frac{1}{2L} \sum_{i}^{L-1} 
\left\{ 1 -\left( \left[ v(0)-u(0)\right] \left[ u^{\dagger}(0) + v^{\dagger}(0) \right] \right)_{i,i+1} \right\} \;.
\end{equation}
%
 
%-------------------------------------------------------------------------------------
\section{Results} \label{results:sec}
%-------------------------------------------------------------------------------------
%
Our results for the dynamics are obtained by integrating numerically the time-dependent
Bogoliubov-de Gennes equations (\ref{BdG_tdep:eqn}). As initial point of the evolution it is 
enough to consider $t_{\rm in}=-5\tau$, taking ${\bf u}_{\mu}^{\rm in}$ and ${\bf v}_{\mu}^{\rm in}$
from the diagonalization of $H(t_{\rm in})$ according to Eq.~(\ref{BdG_static:eqn}): we checked that 
our results do not depend on the precise value of $t_{\rm in}$, as long as it is not too small. 
We considered systems up to $L=512$ and annealing times up to $\tau = 1000$. 
Ensemble averages are calculated over a suitably large number of disorder realizations (of the order of $1000$). 
The analysis of the instantaneous spectrum and its statistics has been obtained by solving the static 
Bogoliubov-de Gennes eigenvalue equations (\ref{BdG_static:eqn}) for systems up to $L=512$.

%.................................................................................
\subsection{Landau-Zener transitions and disorder}
\begin{figure}
\epsfig{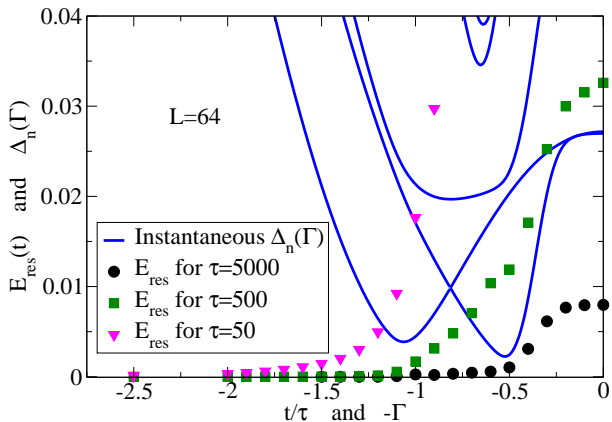}
\caption{(Color online) Residual energy $E_{\rm res}(t)$ versus $t$ for a given instance with $L=64$ of the 
random Ising model with transverse field, at different values of $\tau$. The solid lines
are the lowest-lying instantaneous spectral gaps $\Delta_n$ as a function of $\Gamma$. 
%The red-dashed 
%line is the best fit to the lowest gap used to calculate the Landau-Zener transition 
%rates. For a given value of $\tau$ (different $\tau$ are in different colors) the 
%residual energy obtained by the integration of the equation of motion and the 
%Landau-Zener estimates are represented with squares and down triangles respectively.
}
\label{evolution_L:fig}
\end{figure}

%.................................................................................
%
In order to get an initial understanding on the mechanisms that lead to breaking of 
adiabaticity in the present system, 
%and compare them with the ordered case, 
it is instructive to consider in more detail the time-evolution of a single realization of the disorder. 
To be specific, Fig.~(\ref{evolution_L:fig}) shows the time-evolution of the residual 
% BELOW ADDED FORMULA
energy $E_{\rm res}(t)=\langle\Psi(t)|H(t)|\Psi(t)\rangle-E_{\rm gs}(\Gamma(t))$, 
where $E_{\rm gs}(\Gamma(t))$ is the instantaneous ground state energy corresponding to $\Gamma(t)$, 
for a single $L=64$ sample and for values of $\tau$ up to $5000$. 
We also plot the instantaneous spectral gaps of the problem (thick solid lines)
obtained by diagonalizing the Hamiltonian for any given value of the parameter $\Gamma$.
As mentioned previously, the dynamics conserves the fermion parity, so that only excitations in the same 
fermion parity sector are accessible. If we order the single-particle eigenvalues as 
$\epsilon_1\le \epsilon_2\le \cdots \le \epsilon_L$, then the lowest excited state accessible to 
the dynamics (i.e., conserving the fermionic parity) is associated with an excitation 
energy $\Delta_1=2(\epsilon_1+\epsilon_2)$, rather than $\Delta=2\epsilon_1$.
The next excited state is $\Delta_2=2(\epsilon_1+\epsilon_3)$, and so on. 
These are the instantaneous gaps shown in Fig.~(\ref{evolution_L:fig}). 

An important feature which emerges from this example is that one cannot in general locate 
a single specific value of $\Gamma$ where the minimum and most important gap is present. 
Certainly, typically the first occurrence of a small gap during the annealing trajectory is close 
to the critical point, $\Gamma_c=1$. 
Usually, this critical-point gap is also the smallest one that the systems encounters during its evolution. 
However, it can happen, as Fig.~(\ref{evolution_L:fig}) shows, that the system safely goes through the 
critical-point small gap (see $\tau=5000$ results) but then looses adiabaticity due to a comparable gap 
encountered later on (here at $\Gamma \sim 0.5$). 
Once adiabaticity is lost, the system will generally miss to follow the first
excited state either, getting more and more excited as time goes by. 
%
%---------------------------------------------------------------------------------------------------
%---------------------------------------------------------------------------------------------------
%
%Moreover the adiabaticity can also be lost even at an higher gap. The reason is  that
%what is important in the Landau-Zener mechanism is not only the gap encountered, but also
%the effective speed with which one crosses it, which has to do with the {\em slope} of the
%approaching eigenvalues. 
%More precisely, if we consider a standard $2\times 2$ LZ problem 
%%
%\begin{equation} \label{lz_rot:eqn}
%H_{\rm LZ}(t) = -vt \, \sigma^z - \Delta \sigma^x \;,
%\end{equation}
%%
%where $\Delta$ is the tunneling amplitude between the two states, and $\pm vt$ are 
%the two eigenvalues which in absence of $\Delta$ would cross at $t=0$, the probability 
%of missing
%the ground state and ending up in the excited state at $t=+\infty$ is given by
%the LZ-formula $P_{\rm ex}^{\rm LZ}(t=\infty)=e^{-\pi\Delta^2/\hbar v}$.
%In our problem, however, if we sweep the coupling $\Gamma(t)$ linearly in time, $\Gamma(t)=-t/\tau$,
%and the two crossing eigenvalues $E^0_{\pm}$ are given by $E^0_{\pm}-\tilde{E} = \pm \alpha (\Gamma-\tilde{\Gamma})
%\sim \pm (\alpha/\tau) (t+\tilde{\Gamma}\tau)$, where $\tilde{\Gamma}$ is the value of $\Gamma$ at
%which the crossing occurs and $\alpha$ is the relative slope of two crossing eigenvalues as a function
%of $\Gamma$, then it is quite clear that the ``velocity'' $v$ in the LZ formula should be given
%not just by $1/\tau$, but rather by $\alpha/\tau$: eigenvalues that approach each other in a steeper
%way require, effectively, a small rate $\tau^{-1}$. 

It seems clear that the analysis of the adiabatic dynamics of a disordered
Ising chain requires a knowledge of the statistics of these low-lying gaps in the spectrum 
(in the pertinent parity sector).
We concentrate our attention on the region close to the critical point, where the smallest gaps are found, 
for large $L$.

\begin{figure}
\epsfig{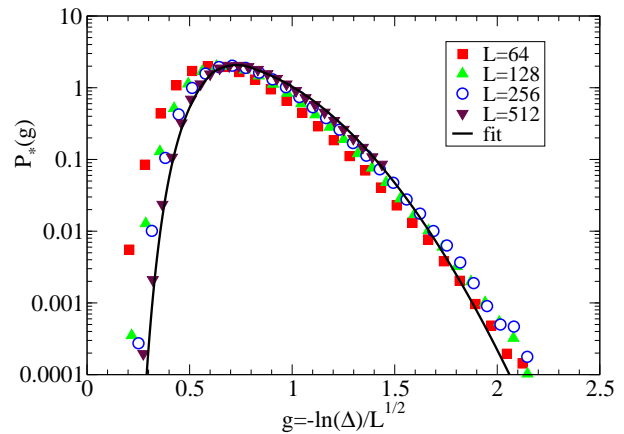}
\caption{(Color online) 
Distribution of $\Delta_1=2(\epsilon_1+\epsilon_2)$, the smallest gap relevant for the dynamics,
at the critical point $\Gamma_c=1$ for different systems sizes, showing the collapse of the distributions
$P(\Delta_1,L)$ when the scaling variable $g=-\log{(\Delta_1})/\sqrt{L}$ is used. 
The resulting distribution is the $P_*(g)$ discussed in the text.
}
\label{gap2:fig}
\end{figure}

We start asking how these smallest gaps are distributed, for different realizations of the disorder.
Let us denote by $P(\Delta_1,L)$ the distribution of gaps $\Delta_1=2(\epsilon_1+\epsilon_2)$ 
(the lowest one relevant for the dynamics) for a chain of length $L$, assumed to be normalized:
$\int_0^{\infty} d\Delta_1 \; P(\Delta_1,L) = 1$. 
For the smallest gap $\Delta=2\epsilon_1$, 
% HERE CHANGED REF and COMMENTED A WHOLE LINE
%--- which, as mentioned above, is not relevant for the dynamics, due to parity conservation ---, 
Young and Rieger \cite{Young_PRB96} have shown that the correct scaling variable which
makes the critical point distribution universal, for different $L$, is $-\log{(\Delta)}/\sqrt{L}$. 
By using a scaling variable of the same form, $g=-\log{(\Delta_1})/\sqrt{L}$, we see that
the gaps $\Delta_1$ are also distributed in the same universal way, see Fig.~(\ref{gap2:fig}).
This implies that at the critical point, $P_*(g)=\sqrt{L}e^{-g\sqrt{L}}P(e^{-g\sqrt{L}};L)$ is, 
for large $L$, universal and normalized. 
As a consequence, gaps at the critical point have an extremely wide distribution, for large $L$, with 
typical gaps which are exponentially small \cite{Fisher_PRB95,Young_PRB96,Fisher_PRB98} in the system size: 
$[\Delta_1]_{\rm typ}\propto e^{-C\sqrt{L}}$.

%
%---------------------------------------------------------------------------------------------------
%---------------------------------------------------------------------------------------------------
%

%.................................................................................
\subsection{Density of kinks}
\begin{figure}
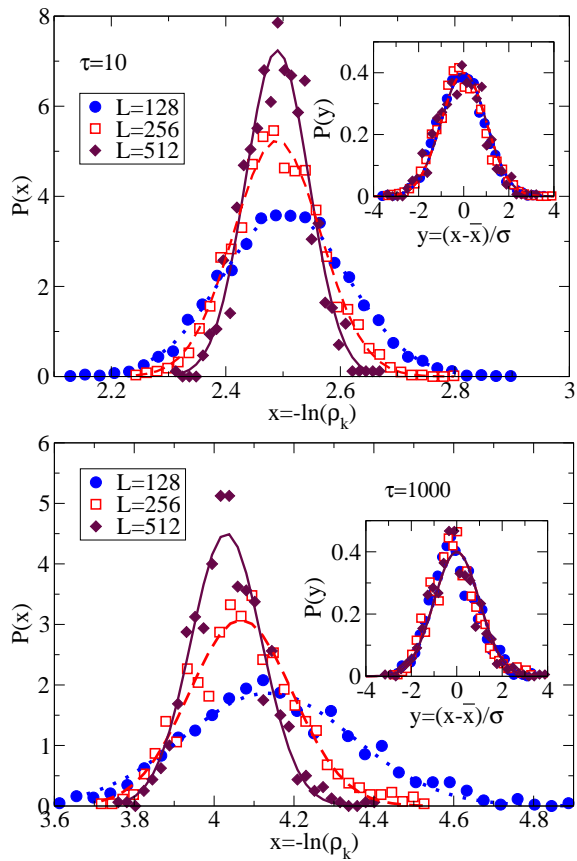

\begin{center}
\epsfig{file=density_distr_t10.eps,width=7.5cm,angle=0,clip=}
\epsfig{file=density_distr_t1000.eps,width=7.5cm,angle=0,clip=}
\end{center}
\caption{(Color online) Probability distribution for the logarithm of the density of defects $x=-\ln \rho_k$, 
for two different annealing rates $\tau$.
The distribution function is universal and log-normal with a variance $\sigma_L$ which scales 
as $1/\sqrt{L}$. In the insets we show the data collapse of all the curves when plotted
as a function of the reduced variable $(x-\bar{x})/\sigma_L$, where $x=-\ln{\rho_k}$.}
\label{d_distrib:fig}
\end{figure}

\begin{figure}
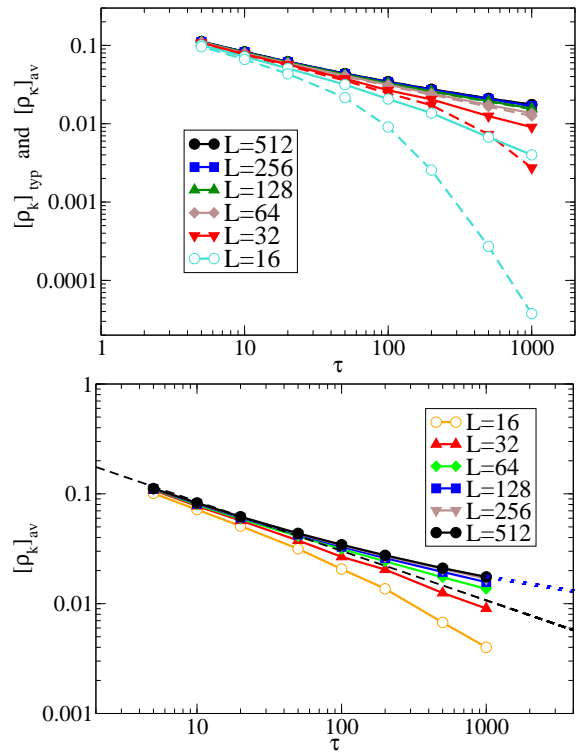

\center
\epsfig{file=global_d_vs_tq.eps,width=7.5cm,angle=0,clip=}
\epsfig{file=d_res.eps,width=7.5cm,angle=0,clip=}
\caption{(Color online) 
Top: Comparison between average $[\rho_k]_{\rm av}$ and typical $[\rho_k]_{\rm typ}=e^{[\ln{\rho_k}]_{\rm av}}$ 
kink density for different system sizes on varying the annealing rate $\tau$. 
The same symbol is used for both cases. 
The typical value (dashed line) lies always below the average value (continuous line), but
the difference between the two is negligible for $L\ge 128$.
Bottom: Average kink density $[\rho_k]_{\rm av}$ 
as a function of the annealing rate $\tau$ for chains of different lengths $L=16,32,64,128,256,512$.
The data for $[\rho_k]_{\rm av}$ are the same appearing in the top part of the figure.
The dashed line is a power-law describing the small-$\tau$ behavior, 
$[\rho_k]_{\rm av}(\tau)\sim \tau^{-0.5}$.
The solid thick line through the $[\rho_k]_{\rm av}$ data is a fit with a function $A/\log^2{(\gamma\tau)}$,
described in the text.
The averages are calculated over $1000$ different realizations of disorder.}
\label{d_res:fig}
\end{figure}

%.................................................................................
%
Given the wide distribution of the instantaneous gaps, it is important to understand how this reflects 
itself in the distribution of various observables. We first consider the behavior of the density 
of defects $\rho_k$ defined in Eq.(\ref{ddefects}). 
The results for the probability distribution function of $\rho_k$, $P(\rho_k)$, are 
presented in Fig.~(\ref{d_distrib:fig}) for $\tau =10$ and $\tau =1000$. 
%
%---------------------------------------------------------------------------------------------------
%---------------------------------------------------------------------------------------------------
%
The distribution $P(\rho_k)$, for given $\tau$, is found to be approximately log-normal: 
\[
P(\rho_k) = \frac{1}{\sqrt{2\pi}\sigma_L} \; \frac{1}{\rho_k} \; 
e^{-\left(\ln{\rho_k} - \overline{\ln{\rho_k}}\right)^2/2\sigma_L^2} \;,
\]
with a standard deviation $\sigma_L$ decreasing as $1/\sqrt{L}$. 
The data collapse of the results for different $L$, in terms of the variable 
$(\ln{\rho_k} - \overline{\ln{\rho_k}})/\sigma_L$, shown in the inset, qualifies 
the accuracy of this statement. 
This $\sqrt{L}$-reduction of the width of the log-normal distribution $P(\rho_k)$ with increasing $L$
is at variance with the result obtained for the distribution of the gaps at the critical point,
whose width {\em increases} as $\sqrt{L}$: here, on the contrary, the correct scaling variable appears to be 
$(\ln{\rho_k} - \overline{\ln{\rho_k}}) \sqrt{L}$, rather than $(\ln{\rho_k} - \overline{\ln{\rho_k}})/\sqrt{L}$. 
This width reduction, for increasing $L$, implies that the {\em average} density of defects $[\rho_k]_{\rm av}$
approaches the {\em typical} value $[{\rho_k}]_{\rm typ}=e^{[\ln{\rho_k}]_{\rm av}}$ for large enough $L$,
since $[\rho_k]_{\rm av} = e^{\overline{\ln{\rho_k}}+\sigma_L^2/2 }$ implies that:
\begin{eqnarray}
\frac{[\rho_k]_{\rm av}-[\rho_k]_{\rm typ}}{[\rho_k]_{\rm typ}} = e^{\sigma_L^2/2}-1 \sim \frac{1}{L} \;.
\end{eqnarray}
This fact is shown explicitly in Fig.(\ref{d_res:fig}) (top), where we see that large deviations between 
$[{\rho_k}]_{\rm typ}=e^{[\ln{\rho_k}]_{\rm av}}$ and $[{\rho_k}]_{\rm av}$ are seen only for $L\le 64$.
For large systems, $L\ge 128$, the two quantities are essentially coincident, for all values of $\tau$.
%
%--------------------------------CHANGED PART
%-------------------------START OLD PART
%
%Despite the universal behavior of the distribution $P(\rho_k)$ at all annealing rates, the 
%behavior of $[\rho_k]_{\rm av}(\tau)$ changes drastically between short and long $\tau$'s \cite{Dziarmaga_PRB06}. 
%Fig.~(\ref{d_res:fig})(bottom) focuses on the average kink density 
%$[\rho_k]_{\rm av}$ for various $L$, as a function of $\tau$. 
%The initial small-$\tau$ behavior of $[\rho_k]_{\rm av}(\tau)$, indicated by
%the dashed line in Fig.~(\ref{d_res:fig}), is power law, $[\rho_k]_{\rm av}(\tau)\sim \tau^{-0.5}$, i.e.,
%exactly what one finds for the ordered Ising chain~\cite{Zurek_PRL05}, where the result is interpreted in terms of the
%the Kibble-Zureck mechanism. 
%
%-------------------------END OLD PART
%-------------------------NEW PART
%
Despite the universal behavior of the distribution $P(\rho_k)$ at all annealing rates, the 
behavior of $[\rho_k]_{\rm av}(\tau)$ changes drastically between short and long $\tau$'s \cite{Dziarmaga_PRB06}. 
Fig.~(\ref{d_res:fig})(bottom) focuses on the average kink density $[\rho_k]_{\rm av}$ for various $L$, as a function of $\tau$. 
The initial small-$\tau$ behavior of $[\rho_k]_{\rm av}(\tau)$, indicated by
the dashed line in Fig.~(\ref{d_res:fig}), seems a power-law, $[\rho_k]_{\rm av}(\tau)\sim \tau^{-0.5}$, i.e.,
exactly what one finds for the ordered Ising chain~\cite{Zurek_PRL05}, where the result is interpreted in terms of the
the Kibble-Zurek mechanism. A possible explanation resides in the fact that our model presents a Griffiths phase 
extending for all $\Gamma > \Gamma_c$~\cite{Igloi_PRB99}.
This phase is characterized by a gap $\Delta \sim L^{-z}$, where the dynamical exponent $z(\Gamma)$ is a 
continuous function of the parameter $\Gamma$, diverging, $z\rightarrow\infty$, for $\Gamma\rightarrow \Gamma_c$,
while saturating to a constant for large $\Gamma$. 
The second gap, which is relevant for our dynamical problem, shows a similar behavior,\cite{Igloi_PRB99} $\Delta_1\sim L^{-z'}$,
with a dynamical exponent $z'(\Gamma)=z(\Gamma)/2$.
For fast annealing rates, the system loses adiabaticity before reaching the critical point, well inside the $\Gamma>\Gamma_c$
Griffiths phase.  
As in the ordered case, the gaps exhibited by such a phase would induce a defect density decreasing as a power-law 
of the annealing time $\tau$, with the crucial difference that the power-law exponent is not constant here, due to the
$\Gamma$-dependence of $z'$. 
One should expect, presumably, a gradual crossover with a power-law exponent which becomes
smaller and smaller, connecting in a gentle way with the large $\tau$ behavior of $[\rho _k]_{av}$, 
which shows marked deviations from a power-law behavior. 
%
%-------------------------END CHANGED PART
%
Dziarmaga, based on scaling arguments~\cite{Dziarmaga_PRB06}
showed that at large $\tau$ the density of kinks should decrease as the inverse
square of the logarithm of $\tau$. Our data for the largest systems agree very well
with this prediction, as the best-fit (solid line in Fig.~(\ref{d_res:fig})) shows.
%
%---------------------------------------------------------------------------------------------------
%---------------------------------------------------------------------------------------------------
%

\begin{figure}[t]
\center
\epsfig{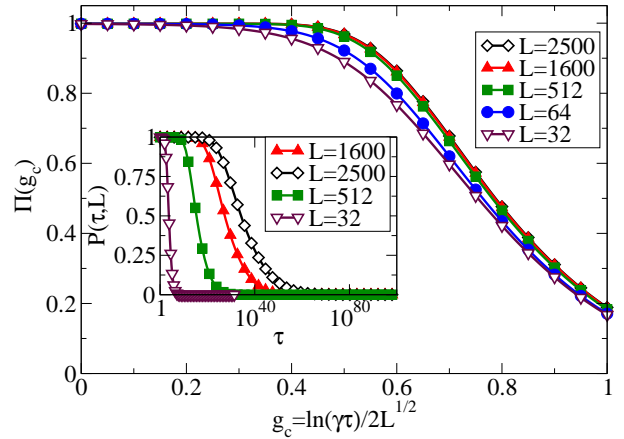}
\caption{(Color online) Approach to the universal function $\Pi(g_c)$ for increasing chain lengths $L$,
see text.
All data from $L\ge 512$ collapse well into a single curve.
Inset: $P_{\rm ex}^{\rm cr.point}(\tau,L)$ obtained from the integral in Eq.~(\ref{P_cr_point:eqn}) 
versus $\tau$ for different values of $L$.
}
\label{Pi_univ:fig}
\end{figure}

A bound to $[\rho_k]_{\rm av}(\tau)$ can also be constructed by a Landau-Zener argument --- complemented 
by a knowledge of the distribution of the first gap $P(\Delta_1,L)$ ---, in a similar fashion to that 
presented by Zurek {\em et al.} \cite{Zurek_PRL05} for the ordered Ising case.
The derivation starts by considering the probability $P_{\rm ex}(\tau,L)$ of loosing adiabaticity 
for a system of size $L$, when turning off $\Gamma$ with an annealing rate $\tau^{-1}$.
Evidently, $P_{\rm ex}(\tau,L) \ge P_{\rm ex}^{\rm cr.point}(\tau,L)$, where we have denoted by
$P_{\rm ex}^{\rm cr.point}(\tau,L)$ the probability of getting excited by Landau-Zener events 
{\em at the critical point} (indeed, we have seen that there is a chance of getting excited also by
gaps well below the critical point). 
$P_{\rm ex}^{\rm cr.point}(\tau,L)$, in turn, can be constructed by knowing the distribution of
the gaps $\Delta_1$ at the critical point, and the simple two-level Landau-Zener formula 
$P_{\rm ex}^{\rm LZ}=e^{-\pi\Delta_1^2\tau/(4\hbar \alpha)}$ ($\alpha$ being the slope of the
two approaching eigenvalues).
% and the factor $4$ in the denominator being due to the fact 
% that the tunneling amplitude is here $\Delta_1/2$). 
Lumping all constants together, $\gamma=\pi/(4\hbar\alpha)$, we write 
$P_{\rm ex}^{\rm LZ}=e^{-\gamma\tau\Delta_1^2}$ and assume that the distribution of 
$\gamma \propto \alpha^{-1}$ is not important in our estimate, while 
that of $\Delta_1$ is, so that:
\begin{eqnarray} \label{P_cr_point:eqn}
P_{\rm ex}^{\rm cr.point}(\tau,L) &=& \int_0^\infty d\Delta_1\; P(\Delta_1,L) \; e^{-\gamma\tau\Delta_1^2} \nonumber \\
&=& \int_{-\infty}^{\infty} dg\; P_{*}(g) \; e^{-\gamma\tau e^{-2\sqrt{L}g}} \;,
\end{eqnarray}
where the second equality follows from switching to the scaling variable $g=-\log{(\Delta_1})/\sqrt{L}$. 
Obviously, for $\tau=0$ we correctly have 
$P_{\rm ex}^{\rm cr.point}(\tau=0,L)=\int_{-\infty}^{\infty} dg\; P_{*}(g)=1$,
from the normalization condition. 
When $\tau$ is finite, the LZ factor $e^{-\gamma\tau e^{-2\sqrt{L}g}}$ provides a lower cut-off in the integral 
at a characteristic $g_c=\log{(\gamma\tau)}/(2\sqrt{L})$, and this cut-off is sharper and sharper as $L$ increases:
one can verify that, for large $L$, 
$e^{-\gamma\tau e^{-2\sqrt{L}g}}\approx \theta(g-g_c)$. 
As a consequence, for large enough $L$ we can rewrite:
\begin{eqnarray} \label{Pi_univ:eqn}
P_{\rm ex}^{\rm cr.point}(\tau,L) \approx \Pi \left(g_c \right)\equiv \int_{g_c}^{\infty} dg\; P_{*}(g) \;,
\end{eqnarray}
i.e., $P_{\rm ex}^{\rm cr.point}(\tau,L)$ turns out to be a universal function of the
scaling variable $g_c=\log{(\gamma\tau)}/(2\sqrt{L})$, for $L$ large. 
This universal function $\Pi(g_c)$ is shown in Fig.(\ref{Pi_univ:fig}), 
where we see that data for $L\ge 512$ collapse into a single curve. 
%
%---------------------------------------------------------------------------------------------------
%---------------------------------------------------------------------------------------------------
The density of kinks for large $\tau$, and large enough $L$, can be obtained by evaluating  
the typical length $\tilde{L}_{\epsilon}(\tau)$ of a defect-free region upon annealing, 
% HERE MODIFIED.
$\epsilon$ being a small quantity of our choice, denoting the probability of getting excited.
Since $P_{\rm ex}^{\rm cr.point}(\tau,L)\approx \Pi(g_c)$ is a {\em lower bound} for
$P_{\rm ex}(\tau,L)$, we have that
\begin{equation} \label{Ltilde:eqn}
\tilde{L}_{\epsilon}(\tau) \le \frac{ \log^2{(\gamma\tau)} }{ [\Pi^{-1}(\epsilon)]^2 } \;,
\end{equation}
where $\Pi^{-1}$ denotes the inverse function of $\Pi$.
If we now identify the inverse of the defect-free region length, $\tilde{L}^{-1}_{\epsilon}(\tau)$,  
with the density of kinks $\rho_{k}(\tau)$, we get the following lower bound for the latter:
\begin{equation} \label{rhok_bound:eqn}
\rho_{k}(\tau) \sim \frac{1}{\tilde{L}_{\epsilon}(\tau)} 
\ge \frac{ [\Pi^{-1}(\epsilon)]^2 }{ \log^2{(\gamma\tau)} } \;.
\end{equation}
On the basis of this argument, we conclude that the density of kinks cannot decrease faster than
$1/\log^2{(\gamma\tau)}$ for large $\tau$, which agrees with the argument discussed by 
Dziarmaga \cite{Dziarmaga_PRB06}.

%....................................................
\subsection{Residual energy}
%....................................................
%
In the ordered case the residual energy per spin is simply proportional to the kink-density,
$E_{\rm res}/L=2J\rho_{k}$, while here, evidently, kinks sitting at small $J_i$'s are favored, on average, by the
adiabatic evolution process. It is therefore of importance to analyze the scaling of the residual 
energy that, as we will show, differs quantitatively from that of the kink density. 
Since kinks will be formed on the weak links, one expects on general grounds that the 
residual energy would decay faster than the kink-density for large $\tau$'s.

\begin{figure}
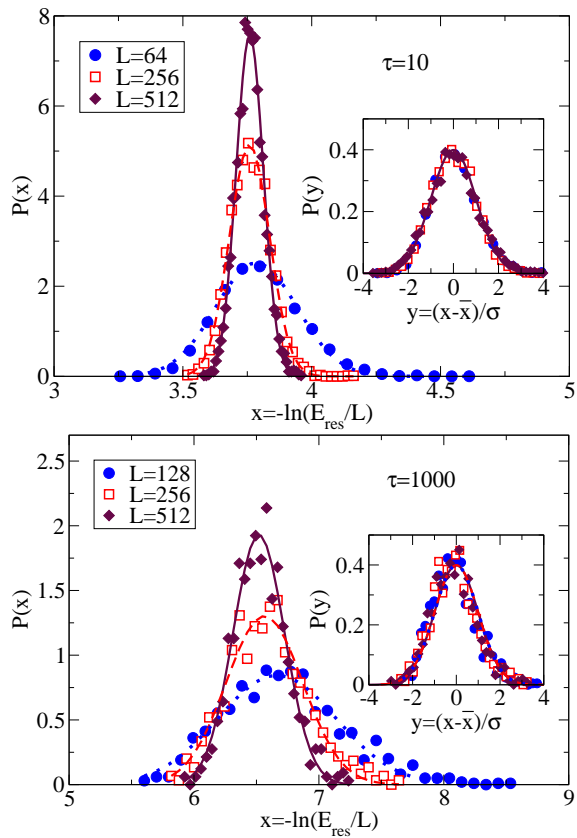

\begin{center}
\epsfig{file=E_res_distr_t10.eps,width=7.5cm,angle=0,clip=}
\epsfig{file=E_res_distr_t1000.eps,width=7.5cm,angle=0,clip=}
\end{center}
\caption{(Color online) 
Probability distribution for the residual energy per site at two different annealing rates $\tau^{-1}$.
The distribution function is universal and log-normal with a variance which scales 
as $1/\sqrt{L}$. In the insets we show the data collapse.}
\label{e_distrib:fig}
\end{figure}

As in the case of the kink density, we first analyze the probability distribution for the residual 
energy per site, which we present in Fig.(\ref{e_distrib:fig}). Once again the residual 
energies are approximately log-normal distributed and can be reduced to a universal form (see the insets)
when properly rescaled, i.e., in terms of the variable 
$(\ln{(E_{\rm res}/L)} - \overline{\ln{(E_{\rm res}/L)}} )\sqrt{L}$.
%
%---------------------------------------------------------------------------------------------------
%---------------------------------------------------------------------------------------------------

The average residual energy per site $[E_{\rm res}/L]_{\rm av}$ as a function of the annealing time
$\tau$ shows a crossover from a power-law decay, 
approximately $\tau^{-1}$ for fast quenches, to a much slower decay 
(see below) for slow evolutions. It is interesting to note that although for fast quenches the disorder
% HERE CHANGED irrelevant --> to play a minor role
is considered to play a minor role,
nevertheless the exponent of the decay of the residual energy differs 
from that of the kink density. The analysis of the regimes of large $\tau$'s is more delicate.
The LZ argument given above tells us nothing about the behavior of the residual energy 
for large $\tau$. We then proceed as follows.
%
%---------------------------------------------------------------------------------------------------
\begin{figure}[tbp]
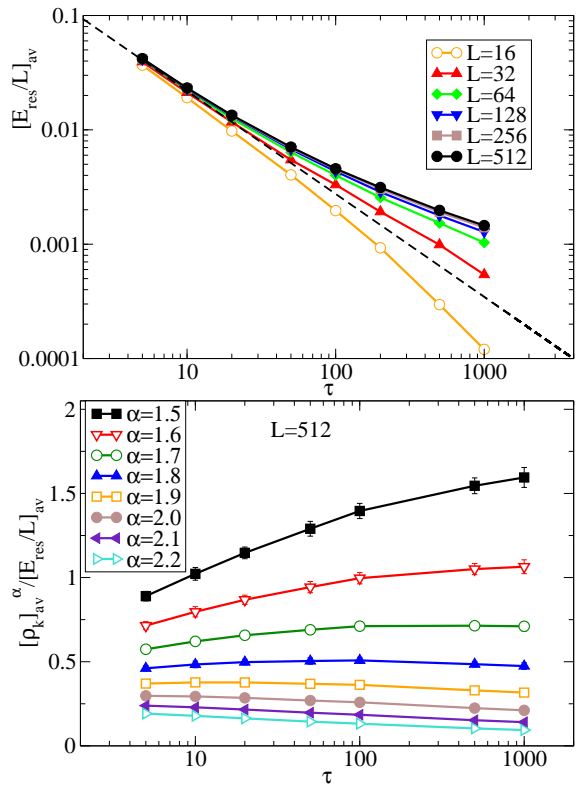

\center
\epsfig{file=e_res.eps,width=7.5cm,angle=0,clip=}
\epsfig{file=rho_over_e_res.eps,width=7.5cm,angle=0,clip=}
\caption{
(Color online) Top: Average residual energy per site $[E_{\rm res}/L]_{\rm av}$ 
as functions of the annealing rate $\tau$ for chains of different lengths $L=16,32,64,128,256,512$.
The dashed line is the power-law describing the small-$\tau$ behavior, 
$[E_{\rm res}/L]_{\rm av}(\tau)\sim \tau^{-1}$.
Averages are calculated over $1000$ realizations of disorder.
Bottom: The ratio of the density of kinks and the residual energy versus $\tau$, used 
to extract the power of the log-dependence of $E_{res}$.
}
\label{e_res:fig}
\end{figure}
%---------------------------------------------------------------------------------------------------
%
Assuming for the residual energy a logarithmic behavior similar to that found for $\rho_k$
\begin{equation}
\left[\frac{E_{\rm res}}{L}\right]_{\rm av} \sim \frac{1}{\log^{\zeta}{(\gamma\tau)}} \;,
\end{equation}
we can determine $\zeta$ from the data of Fig.~(\ref{e_res:fig})(Top) by plotting the ratio of 
$[\rho_{k}]_{\rm av}^{\alpha}$ and $[E_{\rm res}/L]_{\rm av}$ versus $\tau$ for several values of $\alpha$,
as done in Fig.~(\ref{e_res:fig})(Bottom). 
If $[\rho_{k}]_{\rm av}\sim \log^{-2}{(\gamma\tau)}$, then the value of $\alpha$ which makes this ratio 
constant is:
\begin{eqnarray} \label{alpha_value}
\frac{[\rho_{k}]_{\rm av}^{\alpha}}{[E_{\rm res}/L]_{\rm av}} \propto \log^{\zeta-2\alpha}{(\gamma\tau)} 
\sim \mbox{const.} \hspace{1mm} \Longleftrightarrow \hspace{1mm} \alpha=\zeta/2 \;.
\end{eqnarray}
Numerically, see Fig.~(\ref{e_res:fig}), we find $\alpha\approx 1.7\pm 0.1$, which 
implies $\zeta\approx 3.4\pm 0.2$.
  
%--------------------------------------------------------------------------------
\section{Discussion and conclusions} \label{discussion:sec}
%--------------------------------------------------------------------------------
%
In this paper we have studied the adiabatic quantum dynamics of a one-dimensional disordered Ising 
model across its quantum critical point. 
Our main results can be summarized in the dependence of the average kink density 
(see however~\cite{Dziarmaga_PRB06}) and residual energies as a function of the annealing rate
\[
\begin{array}{lll}
\left[\rho_{\rm k}\right]_{\rm av} & \sim \tau^{-0.5}  			& \mbox{fast quenches} \\ 
\left[\rho_{\rm k}\right]_{\rm av} & \sim \left(\log{\tau}\right)^{-2}  	& \mbox{large} \; \tau \;,\\
\end{array} 
\]
\[
\begin{array}{lll}
\left[E_{\rm res}/L\right]_{\rm av} & \sim \tau^{-1}  &  \mbox{fast quenches} \\ 
\left[E_{\rm res}/L\right]_{\rm av} & \sim \left(\log{\tau}\right)^{-\zeta}  & \mbox{large}\; \tau,
\;\; \mbox{with} \; \zeta\sim 3.4 \;.
\end{array} 
\]
Although the dynamics is dominated by a very wide distribution of gaps at the critical point,
% HERE ELIMINATED HYPHENS
$P_*(-\ln{(\Delta_1)}/\sqrt{L})$ (see Fig.~(\ref{gap2:fig})),
we find that the distribution for both these quantities are log-normal but with a variance that decrease, 
like $1/\sqrt{L}$, for increasing chain length $L$: typical and average values, therefore, coincide for large $L$.
The wide distribution of gaps, on the other hand, with its characteristic $\ln{(\Delta_1)}/\sqrt{L}$ scaling,
is responsible, within a Landau-Zener theory, for the extremely slow decay of the average density of
kinks, $[\rho_k]_{\rm av}\sim 1/(\ln{\tau})^2$. 
% HERE ADDED SENTENCE.
This discussion applies only for reasonably large sizes $L$. 
If $L$ is small, the minimum gap $\Delta_1$ of a given instance can be sufficiently large 
that the adiabatic regime, predicted to occur beyond a characteristic $\tau_c \propto \Delta_1^{-2}$, 
is actually seen: a fast decay of $\rho_{\rm k}$ and $E_{\rm res}/L$ 
is expected \cite{Suzuki_JPSJ05} for $\tau > \tau_c$, in such a case. 

It is interesting to compare these results with those of a classical thermal annealing, where,
according to Huse and Fisher \cite{Huse_Fisher_PRL86}, the residual energy also shows a 
logarithmic behavior
\[
E^{\rm CA}_{\rm res}(\tau)/L \sim \left(\log{\tau}\right)^{-\zeta_{CA}}  \hspace{5mm} \zeta_{CA}\le 2 \;,
\]
but with an exponent $\zeta_{CA}$ which is bound by $\zeta_{CA}\le 2$. 

If we look at this problem from the perspective of optimization algorithms, it seems that quantum annealing (QA) 
gives a quantitative improvement over classical annealing for the present system, as is indeed found in other cases 
\cite{Kadowaki_PRE98,Lee_Berne_JPC00,Lee_Berne_JPC01,Santoro_SCI02,Martonak_PRB02,Liu_Berne_JCP03,Martonak_PRE04,Stella_PRB05,Stella_PRB06}, 
but not always (Boolean Satisfiability problems seem to be a test case where QA performs worse than 
classical annealing, see Ref.~\onlinecite{Battaglia_PRE05}). 

In this respect, however, several important issues remain to be clarified. 
First of all, AQC-QA has a large freedom in its construction: the choice of the possible
source of quantum fluctuations \cite{Suzuki_PRE07} --- generally speaking, one can take
$H(t)=H_{\rm fin} + \sum_{\lambda} \Gamma_{\lambda}(t) H_{\lambda}$ ---, 
and the time-dependence of the various $\Gamma_{\lambda}(t)$, which need not be linear in 
time \cite{footnote_gamma,Roland_PRA02}.
Regarding the time dependence of the couplings, we simply note that an optimal choice of the 
``speed'' $\dot{\Gamma}(t)$ with which the critical point is crossed can provide an improvement in
the exponents \cite{Roland_PRA02}, but definitely not change a logarithm into a power-law. 
Regarding the possibility of adding extra kinetic terms to $H(t)$, it is clear that terms like
$-\Gamma_{xy}(t)\sum_{i} J_i \sigma^y_i \sigma^y_{i+1}$ (XY-anisotropy) or similar short range interactions
will not change the universality class of the infinite randomness quantum critical point of the present model
\cite{Fisher_PRB95}. 
Hence, a logarithmically-slow AQC-QA is expected also in more general circumstances, for the present 
one-dimensional model.
We expect this to be a genuine consequence of the randomness present in the problem at hand, which makes the 
adiabatic quantum dynamics intrinsically slow and ineffective in reaching the simple classical ferromagnetic 
ground states \cite{footnote_MoritaProofs,Morita_JPA06}.

This is perhaps to be expected in view of the results of Vidal~\cite{Vidal_PRL03}, who 
showed that problems where the entanglement entropy of a block is bound, can be computed 
classically with a comparable efficiency.
Generically, in disordered one-dimensional system the entanglement entropy grows at most logarithmically 
with the system size at a critical point \cite{Refael_PRL04,Laflorencie_PRB05,DeChiara_JSTAT06},
at this is not enough to substantially change the relative efficiency of quantum versus classical
algorithms.

Therefore, the route to investigate seems to be following: search for models in more then one-dimension,
where the entropy of entanglement grows stronger, which, at the same time, have ``gentle''
enough critical point gap distributions.

Acknowledgments -- We are grateful to E. Tosatti, A. Scardicchio, S. Suzuki, H. Nishimori,
A. Ekert, S. Masida, V. Giovannetti, S. Montangero, J.R. Laguna, G. De Chiara, and W.H. Zurek for discussions. 
This research was partially supported by MIUR-PRIN and EC-Eurosqip. 
The present work has been performed within the ``Quantum Information'' research program 
of the {\it Centro di Ricerca Matematica ``Ennio De Giorgi''} at the Scuola Normale Superiore in Pisa.

%%%%%%%%%%%%%%%%%%%%%%%%%%%%%%%%%%%%%%%%%%%%%%%%%%%%%%%%%%%%%%%%%%%%%%%%%
%                               BIBLIOGRAPHY
%%%%%%%%%%%%%%%%%%%%%%%%%%%%%%%%%%%%%%%%%%%%%%%%%%%%%%%%%%%%%%%%%%%%%%%%%
%\bibliographystyle{apsrev}
%\bibliography{QA}

\begin{thebibliography}{57}
\expandafter\ifx\csname natexlab\endcsname\relax\def\natexlab#1{#1}\fi
\expandafter\ifx\csname bibnamefont\endcsname\relax
  \def\bibnamefont#1{#1}\fi
\expandafter\ifx\csname bibfnamefont\endcsname\relax
  \def\bibfnamefont#1{#1}\fi
\expandafter\ifx\csname citenamefont\endcsname\relax
  \def\citenamefont#1{#1}\fi
\expandafter\ifx\csname url\endcsname\relax
  \def\url#1{\texttt{#1}}\fi
\expandafter\ifx\csname urlprefix\endcsname\relax\def\urlprefix{URL }\fi
\providecommand{\bibinfo}[2]{#2}
\providecommand{\eprint}[2][]{\url{#2}}

\bibitem[{\citenamefont{Farhi et~al.}(2001)\citenamefont{Farhi, Goldstone,
  Gutmann, Lapan, Lundgren, and Preda}}]{Farhi_SCI01}
\bibinfo{author}{\bibfnamefont{E.}~\bibnamefont{Farhi}},
  \bibinfo{author}{\bibfnamefont{J.}~\bibnamefont{Goldstone}},
  \bibinfo{author}{\bibfnamefont{S.}~\bibnamefont{Gutmann}},
  \bibinfo{author}{\bibfnamefont{J.}~\bibnamefont{Lapan}},
  \bibinfo{author}{\bibfnamefont{A.}~\bibnamefont{Lundgren}}, \bibnamefont{and}
  \bibinfo{author}{\bibfnamefont{D.}~\bibnamefont{Preda}},
  \bibinfo{journal}{Science} \textbf{\bibinfo{volume}{292}},
  \bibinfo{pages}{472} (\bibinfo{year}{2001}).

\bibitem[{\citenamefont{Finnila et~al.}(1994)\citenamefont{Finnila, Gomez,
  Sebenik, Stenson, and Doll}}]{Finnila_CPL94}
\bibinfo{author}{\bibfnamefont{A.~B.} \bibnamefont{Finnila}},
  \bibinfo{author}{\bibfnamefont{M.~A.} \bibnamefont{Gomez}},
  \bibinfo{author}{\bibfnamefont{C.}~\bibnamefont{Sebenik}},
  \bibinfo{author}{\bibfnamefont{C.}~\bibnamefont{Stenson}}, \bibnamefont{and}
  \bibinfo{author}{\bibfnamefont{J.~D.} \bibnamefont{Doll}},
  \bibinfo{journal}{Chem. Phys. Lett.} \textbf{\bibinfo{volume}{219}},
  \bibinfo{pages}{343} (\bibinfo{year}{1994}).

\bibitem[{\citenamefont{Kadowaki and Nishimori}(1998)}]{Kadowaki_PRE98}
\bibinfo{author}{\bibfnamefont{T.}~\bibnamefont{Kadowaki}} \bibnamefont{and}
  \bibinfo{author}{\bibfnamefont{H.}~\bibnamefont{Nishimori}},
  \bibinfo{journal}{Phys. Rev. E} \textbf{\bibinfo{volume}{58}},
  \bibinfo{pages}{5355} (\bibinfo{year}{1998}).

\bibitem[{\citenamefont{Brooke et~al.}(1999)\citenamefont{Brooke, Bitko,
  Rosenbaum, and Aeppli}}]{Brooke_SCI99}
\bibinfo{author}{\bibfnamefont{J.}~\bibnamefont{Brooke}},
  \bibinfo{author}{\bibfnamefont{D.}~\bibnamefont{Bitko}},
  \bibinfo{author}{\bibfnamefont{T.~F.} \bibnamefont{Rosenbaum}},
  \bibnamefont{and} \bibinfo{author}{\bibfnamefont{G.}~\bibnamefont{Aeppli}},
  \bibinfo{journal}{Science} \textbf{\bibinfo{volume}{284}},
  \bibinfo{pages}{779} (\bibinfo{year}{1999}).

\bibitem[{\citenamefont{Santoro et~al.}(2002)\citenamefont{Santoro,
  {Marto\v{n}\'{a}k}, Tosatti, and Car}}]{Santoro_SCI02}
\bibinfo{author}{\bibfnamefont{G.~E.} \bibnamefont{Santoro}},
  \bibinfo{author}{\bibfnamefont{R.}~\bibnamefont{{Marto\v{n}\'{a}k}}},
  \bibinfo{author}{\bibfnamefont{E.}~\bibnamefont{Tosatti}}, \bibnamefont{and}
  \bibinfo{author}{\bibfnamefont{R.}~\bibnamefont{Car}},
  \bibinfo{journal}{Science} \textbf{\bibinfo{volume}{295}},
  \bibinfo{pages}{2427} (\bibinfo{year}{2002}).

\bibitem[{\citenamefont{Das and Chakrabarti}(2005)}]{Das_Chakrabarti:book}
\bibinfo{author}{\bibfnamefont{A.}~\bibnamefont{Das}} \bibnamefont{and}
  \bibinfo{author}{\bibfnamefont{B.~K.} \bibnamefont{Chakrabarti}},
  \emph{\bibinfo{title}{{Quantum Annealing and Related Optimization Methods}}},
  Lecture Notes in Physics (\bibinfo{publisher}{Springer-Verlag},
  \bibinfo{year}{2005}).

\bibitem[{\citenamefont{Santoro and Tosatti}(2006)}]{Santoro_JPA:review}
\bibinfo{author}{\bibfnamefont{G.~E.} \bibnamefont{Santoro}} \bibnamefont{and}
  \bibinfo{author}{\bibfnamefont{E.}~\bibnamefont{Tosatti}},
  \bibinfo{journal}{J. Phys. A: Math. Gen.} \textbf{\bibinfo{volume}{39}},
  \bibinfo{pages}{R393} (\bibinfo{year}{2006}).

\bibitem[{\citenamefont{Zurek et~al.}(2005)\citenamefont{Zurek, Dorner, and
  Zoller}}]{Zurek_PRL05}
\bibinfo{author}{\bibfnamefont{W.~H.} \bibnamefont{Zurek}},
  \bibinfo{author}{\bibfnamefont{U.}~\bibnamefont{Dorner}}, \bibnamefont{and}
  \bibinfo{author}{\bibfnamefont{P.}~\bibnamefont{Zoller}},
  \bibinfo{journal}{Phys. Rev. Lett.} \textbf{\bibinfo{volume}{95}},
  \bibinfo{pages}{105701} (\bibinfo{year}{2005}).

\bibitem[{\citenamefont{Nielsen and Chuang}(2000)}]{Nielsen_Chuang:book}
\bibinfo{author}{\bibfnamefont{M.}~\bibnamefont{Nielsen}} \bibnamefont{and}
  \bibinfo{author}{\bibfnamefont{I.~L.} \bibnamefont{Chuang}},
  \emph{\bibinfo{title}{Quantum Computation and Quantum Information}}
  (\bibinfo{publisher}{Cambridge University Press}, \bibinfo{year}{2000}).

\bibitem[{\citenamefont{Aharonov et~al.}(2004)\citenamefont{Aharonov, {van
  Dam}, Kempe, Landau, Lloyd, and Regev}}]{Aharonov:proceeding}
\bibinfo{author}{\bibfnamefont{D.}~\bibnamefont{Aharonov}},
  \bibinfo{author}{\bibfnamefont{W.}~\bibnamefont{{van Dam}}},
  \bibinfo{author}{\bibfnamefont{J.}~\bibnamefont{Kempe}},
  \bibinfo{author}{\bibfnamefont{Z.}~\bibnamefont{Landau}},
  \bibinfo{author}{\bibfnamefont{S.}~\bibnamefont{Lloyd}}, \bibnamefont{and}
  \bibinfo{author}{\bibfnamefont{O.}~\bibnamefont{Regev}}, in
  \emph{\bibinfo{booktitle}{Proceedings of the 45th Annual IEEE Symposium of
  Foundations of Computer Science (FOCS'04)}} (\bibinfo{year}{2004}),
  p.~\bibinfo{pages}{42}.

\bibitem[{\citenamefont{Aharonov et~al.}()\citenamefont{Aharonov, {van Dam},
  Kempe, Landau, Lloyd, and Regev}}]{Aharonov_QP04:preprint}
\bibinfo{author}{\bibfnamefont{D.}~\bibnamefont{Aharonov}},
  \bibinfo{author}{\bibfnamefont{W.}~\bibnamefont{{van Dam}}},
  \bibinfo{author}{\bibfnamefont{J.}~\bibnamefont{Kempe}},
  \bibinfo{author}{\bibfnamefont{Z.}~\bibnamefont{Landau}},
  \bibinfo{author}{\bibfnamefont{S.}~\bibnamefont{Lloyd}}, \bibnamefont{and}
  \bibinfo{author}{\bibfnamefont{O.}~\bibnamefont{Regev}},
  \eprint{quant-ph/0405098}.

\bibitem[{foo({\natexlab{a}})}]{footnote_gamma}
\bibinfo{note}{In order to optimize the adiabatic algorithm one should also
  find the optimal time dependence of the coupling constant $\Gamma(t)$, see
  Ref.~\onlinecite{Roland_PRA02}.}

\bibitem[{\citenamefont{Roland and Cerf}(2002)}]{Roland_PRA02}
\bibinfo{author}{\bibfnamefont{J.}~\bibnamefont{Roland}} \bibnamefont{and}
  \bibinfo{author}{\bibfnamefont{N.~J.} \bibnamefont{Cerf}},
  \bibinfo{journal}{Phys. Rev. A} \textbf{\bibinfo{volume}{65}},
  \bibinfo{pages}{042308} (\bibinfo{year}{2002}).

\bibitem[{\citenamefont{Kibble}(1980)}]{Kibble:review}
\bibinfo{author}{\bibfnamefont{T.~W.~B.} \bibnamefont{Kibble}},
  \bibinfo{journal}{Phys. Rep.} \textbf{\bibinfo{volume}{67}},
  \bibinfo{pages}{183} (\bibinfo{year}{1980}).

\bibitem[{\citenamefont{Zurek}(1996)}]{Zurek:review}
\bibinfo{author}{\bibfnamefont{W.~H.} \bibnamefont{Zurek}},
  \bibinfo{journal}{Phys. Rep.} \textbf{\bibinfo{volume}{276}},
  \bibinfo{pages}{177} (\bibinfo{year}{1996}).

\bibitem[{\citenamefont{Bauerle et~al.}(1996)\citenamefont{Bauerle, Bunkov,
  Fisher, Godfrin, and Pickett}}]{Bauerle_NAT96}
\bibinfo{author}{\bibfnamefont{C.}~\bibnamefont{Bauerle}},
  \bibinfo{author}{\bibfnamefont{Y.~M.} \bibnamefont{Bunkov}},
  \bibinfo{author}{\bibfnamefont{S.~N.} \bibnamefont{Fisher}},
  \bibinfo{author}{\bibfnamefont{H.}~\bibnamefont{Godfrin}}, \bibnamefont{and}
  \bibinfo{author}{\bibfnamefont{G.~R.} \bibnamefont{Pickett}},
  \bibinfo{journal}{Nature} \textbf{\bibinfo{volume}{382}},
  \bibinfo{pages}{332} (\bibinfo{year}{1996}).

\bibitem[{\citenamefont{Ruutu et~al.}(1996)\citenamefont{Ruutu, Eltsov, Gill,
  Kibble, Krusius, Makhlin, Placais, Volovik, and Xu}}]{Ruutu_NAT96}
\bibinfo{author}{\bibfnamefont{V.~M.~H.} \bibnamefont{Ruutu}},
  \bibinfo{author}{\bibfnamefont{V.~B.} \bibnamefont{Eltsov}},
  \bibinfo{author}{\bibfnamefont{A.~J.} \bibnamefont{Gill}},
  \bibinfo{author}{\bibfnamefont{T.~W.~B.} \bibnamefont{Kibble}},
  \bibinfo{author}{\bibfnamefont{M.}~\bibnamefont{Krusius}},
  \bibinfo{author}{\bibfnamefont{Y.~G.} \bibnamefont{Makhlin}},
  \bibinfo{author}{\bibfnamefont{B.}~\bibnamefont{Placais}},
  \bibinfo{author}{\bibfnamefont{G.~E.} \bibnamefont{Volovik}},
  \bibnamefont{and} \bibinfo{author}{\bibfnamefont{W.}~\bibnamefont{Xu}},
  \bibinfo{journal}{Nature} \textbf{\bibinfo{volume}{382}},
  \bibinfo{pages}{334} (\bibinfo{year}{1996}).

\bibitem[{\citenamefont{Polkovnikov}(2005)}]{Polkovnikov_PRB05}
\bibinfo{author}{\bibfnamefont{A.}~\bibnamefont{Polkovnikov}},
  \bibinfo{journal}{Phys. Rev. B} \textbf{\bibinfo{volume}{72}},
  \bibinfo{pages}{161201(R)} (\bibinfo{year}{2005}).

\bibitem[{foo({\natexlab{b}})}]{footnote_Polkovnikov}
\bibinfo{note}{For a general approach see also
  Ref.~\onlinecite{Polkovnikov_07:preprint}.}

\bibitem[{\citenamefont{Polkovnikov and
  Gritsev}(2007)}]{Polkovnikov_07:preprint}
\bibinfo{author}{\bibfnamefont{A.}~\bibnamefont{Polkovnikov}} \bibnamefont{and}
  \bibinfo{author}{\bibfnamefont{V.}~\bibnamefont{Gritsev}}
  (\bibinfo{year}{2007}), \eprint{arXiv:0706.0212}.

\bibitem[{\citenamefont{Dziarmaga}(2005)}]{Dziarmaga_PRL05}
\bibinfo{author}{\bibfnamefont{J.}~\bibnamefont{Dziarmaga}},
  \bibinfo{journal}{Phys. Rev. Lett.} \textbf{\bibinfo{volume}{95}},
  \bibinfo{pages}{245701} (\bibinfo{year}{2005}).

\bibitem[{\citenamefont{Pfeuty}(1970)}]{Pfeuty_AP70}
\bibinfo{author}{\bibfnamefont{P.}~\bibnamefont{Pfeuty}},
  \bibinfo{journal}{Ann. Phys. (N.Y.)} \textbf{\bibinfo{volume}{57}},
  \bibinfo{pages}{79} (\bibinfo{year}{1970}).

\bibitem[{\citenamefont{Damski}(2005)}]{Damski_PRL05}
\bibinfo{author}{\bibfnamefont{B.}~\bibnamefont{Damski}},
  \bibinfo{journal}{Phys. Rev. Lett.} \textbf{\bibinfo{volume}{95}},
  \bibinfo{pages}{035701} (\bibinfo{year}{2005}).

\bibitem[{\citenamefont{Damski and Zurek}(2006)}]{Damski_PRA06}
\bibinfo{author}{\bibfnamefont{B.}~\bibnamefont{Damski}} \bibnamefont{and}
  \bibinfo{author}{\bibfnamefont{W.~H.} \bibnamefont{Zurek}},
  \bibinfo{journal}{Phys. Rev. A} \textbf{\bibinfo{volume}{73}},
  \bibinfo{pages}{063405} (\bibinfo{year}{2006}).

\bibitem[{\citenamefont{Cherng and Levitov}(2006)}]{Cherng_PRA06}
\bibinfo{author}{\bibfnamefont{R.~W.} \bibnamefont{Cherng}} \bibnamefont{and}
  \bibinfo{author}{\bibfnamefont{L.}~\bibnamefont{Levitov}},
  \bibinfo{journal}{Phys. Rev. A} \textbf{\bibinfo{volume}{73}},
  \bibinfo{pages}{043614} (\bibinfo{year}{2006}).

\bibitem[{\citenamefont{Fubini et~al.}()\citenamefont{Fubini, Osterloh, and
  Falci}}]{Fubini_07:preprint}
\bibinfo{author}{\bibfnamefont{A.}~\bibnamefont{Fubini}},
  \bibinfo{author}{\bibfnamefont{A.}~\bibnamefont{Osterloh}}, \bibnamefont{and}
  \bibinfo{author}{\bibfnamefont{G.}~\bibnamefont{Falci}}, \bibinfo{note}{to
  appear in New. J. Phys.}, \eprint{cond-mat/0702014}.

\bibitem[{\citenamefont{Schutzhold et~al.}(2006)\citenamefont{Schutzhold,
  Uhlmann, Xu, and Fischer}}]{Schutzhold_PRL06}
\bibinfo{author}{\bibfnamefont{R.}~\bibnamefont{Schutzhold}},
  \bibinfo{author}{\bibfnamefont{M.}~\bibnamefont{Uhlmann}},
  \bibinfo{author}{\bibfnamefont{Y.}~\bibnamefont{Xu}}, \bibnamefont{and}
  \bibinfo{author}{\bibfnamefont{U.~R.} \bibnamefont{Fischer}},
  \bibinfo{journal}{Phys. Rev. Lett.} \textbf{\bibinfo{volume}{97}},
  \bibinfo{pages}{200601} (\bibinfo{year}{2006}).

\bibitem[{\citenamefont{Cucchietti et~al.}(2007)\citenamefont{Cucchietti,
  Damski, Dziarmaga, and Zurek}}]{Cucchietti_PRA07}
\bibinfo{author}{\bibfnamefont{F.~M.} \bibnamefont{Cucchietti}},
  \bibinfo{author}{\bibfnamefont{B.}~\bibnamefont{Damski}},
  \bibinfo{author}{\bibfnamefont{J.}~\bibnamefont{Dziarmaga}},
  \bibnamefont{and} \bibinfo{author}{\bibfnamefont{W.~H.} \bibnamefont{Zurek}},
  \bibinfo{journal}{Phys. Rev. A} \textbf{\bibinfo{volume}{75}},
  \bibinfo{pages}{023603} (\bibinfo{year}{2007}).

\bibitem[{\citenamefont{Latorre and Orus}(2004)}]{Latorre_PRA04}
\bibinfo{author}{\bibfnamefont{J.}~\bibnamefont{Latorre}} \bibnamefont{and}
  \bibinfo{author}{\bibfnamefont{R.}~\bibnamefont{Orus}},
  \bibinfo{journal}{Phys. Rev. A} \textbf{\bibinfo{volume}{69}},
  \bibinfo{pages}{062302} (\bibinfo{year}{2004}).

\bibitem[{\citenamefont{Cincio et~al.}(2007)\citenamefont{Cincio, Dziarmaga,
  Rams, and Zurek}}]{Cincio_PRA07}
\bibinfo{author}{\bibfnamefont{L.}~\bibnamefont{Cincio}},
  \bibinfo{author}{\bibfnamefont{J.}~\bibnamefont{Dziarmaga}},
  \bibinfo{author}{\bibfnamefont{M.~M.} \bibnamefont{Rams}}, \bibnamefont{and}
  \bibinfo{author}{\bibfnamefont{W.~H.} \bibnamefont{Zurek}},
  \bibinfo{journal}{Phys. Rev. A} \textbf{\bibinfo{volume}{75}},
  \bibinfo{pages}{052321} (\bibinfo{year}{2007}).

\bibitem[{\citenamefont{Hopfield and Tank}(1986)}]{Hopfield_SCI86}
\bibinfo{author}{\bibfnamefont{J.~J.} \bibnamefont{Hopfield}} \bibnamefont{and}
  \bibinfo{author}{\bibfnamefont{D.~W.} \bibnamefont{Tank}},
  \bibinfo{journal}{Science} \textbf{\bibinfo{volume}{233}},
  \bibinfo{pages}{625} (\bibinfo{year}{1986}).

\bibitem[{\citenamefont{{M\'ezard} et~al.}(2002)\citenamefont{{M\'ezard},
  Parisi, and Zecchina}}]{Mezard_SCI02}
\bibinfo{author}{\bibfnamefont{M.}~\bibnamefont{{M\'ezard}}},
  \bibinfo{author}{\bibfnamefont{G.}~\bibnamefont{Parisi}}, \bibnamefont{and}
  \bibinfo{author}{\bibfnamefont{R.}~\bibnamefont{Zecchina}},
  \bibinfo{journal}{Science} \textbf{\bibinfo{volume}{297}},
  \bibinfo{pages}{812} (\bibinfo{year}{2002}).

\bibitem[{\citenamefont{Dziarmaga}(2006)}]{Dziarmaga_PRB06}
\bibinfo{author}{\bibfnamefont{J.}~\bibnamefont{Dziarmaga}},
  \bibinfo{journal}{Phys. Rev. B} \textbf{\bibinfo{volume}{74}},
  \bibinfo{pages}{064416} (\bibinfo{year}{2006}).

\bibitem[{\citenamefont{Fisher}(1995)}]{Fisher_PRB95}
\bibinfo{author}{\bibfnamefont{D.~S.} \bibnamefont{Fisher}},
  \bibinfo{journal}{Phys. Rev. B} \textbf{\bibinfo{volume}{51}},
  \bibinfo{pages}{6411} (\bibinfo{year}{1995}).

\bibitem[{\citenamefont{Lieb et~al.}(1961)\citenamefont{Lieb, Schultz, and
  Mattis}}]{Lieb_AP61}
\bibinfo{author}{\bibfnamefont{E.}~\bibnamefont{Lieb}},
  \bibinfo{author}{\bibfnamefont{T.}~\bibnamefont{Schultz}}, \bibnamefont{and}
  \bibinfo{author}{\bibfnamefont{D.}~\bibnamefont{Mattis}},
  \bibinfo{journal}{Ann. Phys. (N.Y.)} \textbf{\bibinfo{volume}{16}},
  \bibinfo{pages}{407} (\bibinfo{year}{1961}).

\bibitem[{\citenamefont{Young}(1997)}]{Young_PRB97}
\bibinfo{author}{\bibfnamefont{A.}~\bibnamefont{Young}},
  \bibinfo{journal}{Phys. Rev. B} \textbf{\bibinfo{volume}{56}},
  \bibinfo{pages}{11691} (\bibinfo{year}{1997}).

\bibitem[{\citenamefont{Young and Rieger}(1996)}]{Young_PRB96}
\bibinfo{author}{\bibfnamefont{A.~P.} \bibnamefont{Young}} \bibnamefont{and}
  \bibinfo{author}{\bibfnamefont{H.}~\bibnamefont{Rieger}},
  \bibinfo{journal}{Phys. Rev. B} \textbf{\bibinfo{volume}{53}},
  \bibinfo{pages}{8486} (\bibinfo{year}{1996}).

\bibitem[{\citenamefont{Fisher and Young}(1998)}]{Fisher_PRB98}
\bibinfo{author}{\bibfnamefont{D.~S.} \bibnamefont{Fisher}} \bibnamefont{and}
  \bibinfo{author}{\bibfnamefont{A.~P.} \bibnamefont{Young}},
  \bibinfo{journal}{Phys. Rev. B} \textbf{\bibinfo{volume}{58}},
  \bibinfo{pages}{9131} (\bibinfo{year}{1998}).

\bibitem[{\citenamefont{Barouch et~al.}(1970)\citenamefont{Barouch, McCoy, and
  Dresden}}]{Barouch_PRA70}
\bibinfo{author}{\bibfnamefont{E.}~\bibnamefont{Barouch}},
  \bibinfo{author}{\bibfnamefont{B.~M.} \bibnamefont{McCoy}}, \bibnamefont{and}
  \bibinfo{author}{\bibfnamefont{M.}~\bibnamefont{Dresden}},
  \bibinfo{journal}{Phys. Rev. A} \textbf{\bibinfo{volume}{2}},
  \bibinfo{pages}{1075} (\bibinfo{year}{1970}).

\bibitem[{\citenamefont{Igloi et~al.}(1999)\citenamefont{Igloi, Juhasz, and
  Rieger}}]{Igloi_PRB99}
\bibinfo{author}{\bibfnamefont{F.}~\bibnamefont{Igloi}},
  \bibinfo{author}{\bibfnamefont{R.}~\bibnamefont{Juhasz}}, \bibnamefont{and}
  \bibinfo{author}{\bibfnamefont{H.}~\bibnamefont{Rieger}},
  \bibinfo{journal}{Phys. Rev. B} \textbf{\bibinfo{volume}{59}},
  \bibinfo{pages}{11308} (\bibinfo{year}{1999}).

\bibitem[{\citenamefont{Suzuki and Okada}(2005)}]{Suzuki_JPSJ05}
\bibinfo{author}{\bibfnamefont{S.}~\bibnamefont{Suzuki}} \bibnamefont{and}
  \bibinfo{author}{\bibfnamefont{M.}~\bibnamefont{Okada}}, \bibinfo{journal}{J.
  Phys. Soc. Jpn.} \textbf{\bibinfo{volume}{74}}, \bibinfo{pages}{1649}
  (\bibinfo{year}{2005}).

\bibitem[{\citenamefont{Huse and Fisher}(1986)}]{Huse_Fisher_PRL86}
\bibinfo{author}{\bibfnamefont{D.~A.} \bibnamefont{Huse}} \bibnamefont{and}
  \bibinfo{author}{\bibfnamefont{D.~S.} \bibnamefont{Fisher}},
  \bibinfo{journal}{Phys. Rev. Lett.} \textbf{\bibinfo{volume}{57}},
  \bibinfo{pages}{2203} (\bibinfo{year}{1986}).

\bibitem[{\citenamefont{Lee and Berne}(2000)}]{Lee_Berne_JPC00}
\bibinfo{author}{\bibfnamefont{Y.~H.} \bibnamefont{Lee}} \bibnamefont{and}
  \bibinfo{author}{\bibfnamefont{B.~J.} \bibnamefont{Berne}},
  \bibinfo{journal}{J. Phys. Chem. A} \textbf{\bibinfo{volume}{104}},
  \bibinfo{pages}{86} (\bibinfo{year}{2000}).

\bibitem[{\citenamefont{Lee and Berne}(2001)}]{Lee_Berne_JPC01}
\bibinfo{author}{\bibfnamefont{Y.~H.} \bibnamefont{Lee}} \bibnamefont{and}
  \bibinfo{author}{\bibfnamefont{B.~J.} \bibnamefont{Berne}},
  \bibinfo{journal}{J. Phys. Chem. A} \textbf{\bibinfo{volume}{105}},
  \bibinfo{pages}{459} (\bibinfo{year}{2001}).

\bibitem[{\citenamefont{{Marto\v{n}\'{a}k}
  et~al.}(2002)\citenamefont{{Marto\v{n}\'{a}k}, Santoro, and
  Tosatti}}]{Martonak_PRB02}
\bibinfo{author}{\bibfnamefont{R.}~\bibnamefont{{Marto\v{n}\'{a}k}}},
  \bibinfo{author}{\bibfnamefont{G.~E.} \bibnamefont{Santoro}},
  \bibnamefont{and} \bibinfo{author}{\bibfnamefont{E.}~\bibnamefont{Tosatti}},
  \bibinfo{journal}{Phys. Rev. B} \textbf{\bibinfo{volume}{66}},
  \bibinfo{pages}{094203} (\bibinfo{year}{2002}).

\bibitem[{\citenamefont{Liu and Berne}(2003)}]{Liu_Berne_JCP03}
\bibinfo{author}{\bibfnamefont{P.}~\bibnamefont{Liu}} \bibnamefont{and}
  \bibinfo{author}{\bibfnamefont{B.~J.} \bibnamefont{Berne}},
  \bibinfo{journal}{J. Chem. Phys.} \textbf{\bibinfo{volume}{118}},
  \bibinfo{pages}{2999} (\bibinfo{year}{2003}).

\bibitem[{\citenamefont{{Marto\v{n}\'{a}k}
  et~al.}(2004)\citenamefont{{Marto\v{n}\'{a}k}, Santoro, and
  Tosatti}}]{Martonak_PRE04}
\bibinfo{author}{\bibfnamefont{R.}~\bibnamefont{{Marto\v{n}\'{a}k}}},
  \bibinfo{author}{\bibfnamefont{G.~E.} \bibnamefont{Santoro}},
  \bibnamefont{and} \bibinfo{author}{\bibfnamefont{E.}~\bibnamefont{Tosatti}},
  \bibinfo{journal}{Phys. Rev. E} \textbf{\bibinfo{volume}{70}},
  \bibinfo{pages}{057701} (\bibinfo{year}{2004}).

\bibitem[{\citenamefont{Stella et~al.}(2005)\citenamefont{Stella, Santoro, and
  Tosatti}}]{Stella_PRB05}
\bibinfo{author}{\bibfnamefont{L.}~\bibnamefont{Stella}},
  \bibinfo{author}{\bibfnamefont{G.~E.} \bibnamefont{Santoro}},
  \bibnamefont{and} \bibinfo{author}{\bibfnamefont{E.}~\bibnamefont{Tosatti}},
  \bibinfo{journal}{Phys. Rev. B} \textbf{\bibinfo{volume}{72}},
  \bibinfo{pages}{014303} (\bibinfo{year}{2005}).

\bibitem[{\citenamefont{Stella et~al.}(2006)\citenamefont{Stella, Santoro, and
  Tosatti}}]{Stella_PRB06}
\bibinfo{author}{\bibfnamefont{L.}~\bibnamefont{Stella}},
  \bibinfo{author}{\bibfnamefont{G.~E.} \bibnamefont{Santoro}},
  \bibnamefont{and} \bibinfo{author}{\bibfnamefont{E.}~\bibnamefont{Tosatti}},
  \bibinfo{journal}{Phys. Rev. B} \textbf{\bibinfo{volume}{73}},
  \bibinfo{pages}{144302} (\bibinfo{year}{2006}), \eprint{cond-mat/0512064}.

\bibitem[{\citenamefont{Battaglia et~al.}(2005)\citenamefont{Battaglia,
  Santoro, and Tosatti}}]{Battaglia_PRE05}
\bibinfo{author}{\bibfnamefont{D.~A.} \bibnamefont{Battaglia}},
  \bibinfo{author}{\bibfnamefont{G.~E.} \bibnamefont{Santoro}},
  \bibnamefont{and} \bibinfo{author}{\bibfnamefont{E.}~\bibnamefont{Tosatti}},
  \bibinfo{journal}{Phys. Rev. E} \textbf{\bibinfo{volume}{71}},
  \bibinfo{pages}{066707} (\bibinfo{year}{2005}).

\bibitem[{\citenamefont{Suzuki et~al.}(2007)\citenamefont{Suzuki, Nishimori,
  and Suzuki}}]{Suzuki_PRE07}
\bibinfo{author}{\bibfnamefont{S.}~\bibnamefont{Suzuki}},
  \bibinfo{author}{\bibfnamefont{H.}~\bibnamefont{Nishimori}},
  \bibnamefont{and} \bibinfo{author}{\bibfnamefont{M.}~\bibnamefont{Suzuki}},
  \bibinfo{journal}{Phys. Rev. E} \textbf{\bibinfo{volume}{75}},
  \bibinfo{pages}{051112} (\bibinfo{year}{2007}).

\bibitem[{foo({\natexlab{c}})}]{footnote_MoritaProofs}
\bibinfo{note}{We mention, however, that for finite-size Ising systems,
  convergence bound have been proved for AQC-QA in terms of power-law annealing
  schedules, see Ref.~\onlinecite{Morita_JPA06}.}

\bibitem[{\citenamefont{Morita and Nishimori}(2006)}]{Morita_JPA06}
\bibinfo{author}{\bibfnamefont{S.}~\bibnamefont{Morita}} \bibnamefont{and}
  \bibinfo{author}{\bibfnamefont{H.}~\bibnamefont{Nishimori}},
  \bibinfo{journal}{J. Phys. A} \textbf{\bibinfo{volume}{39}},
  \bibinfo{pages}{13903} (\bibinfo{year}{2006}).

\bibitem[{\citenamefont{Vidal}(2003)}]{Vidal_PRL03}
\bibinfo{author}{\bibfnamefont{G.}~\bibnamefont{Vidal}},
  \bibinfo{journal}{Phys. Rev. Lett.} \textbf{\bibinfo{volume}{91}},
  \bibinfo{pages}{147902} (\bibinfo{year}{2003}).

\bibitem[{\citenamefont{Refael and Moore}(2004)}]{Refael_PRL04}
\bibinfo{author}{\bibfnamefont{G.}~\bibnamefont{Refael}} \bibnamefont{and}
  \bibinfo{author}{\bibfnamefont{J.}~\bibnamefont{Moore}},
  \bibinfo{journal}{Phys. Rev. Lett.} \textbf{\bibinfo{volume}{93}},
  \bibinfo{pages}{260602} (\bibinfo{year}{2004}).

\bibitem[{\citenamefont{Laflorencie}(2005)}]{Laflorencie_PRB05}
\bibinfo{author}{\bibfnamefont{N.}~\bibnamefont{Laflorencie}},
  \bibinfo{journal}{Phys. Rev. B} \textbf{\bibinfo{volume}{72}},
  \bibinfo{pages}{140408(R)} (\bibinfo{year}{2005}).

\bibitem[{\citenamefont{{De Chiara} et~al.}(2006)\citenamefont{{De Chiara},
  Montanegro, Calabrese, and Fazio}}]{DeChiara_JSTAT06}
\bibinfo{author}{\bibfnamefont{G.}~\bibnamefont{{De Chiara}}},
  \bibinfo{author}{\bibfnamefont{S.}~\bibnamefont{Montanegro}},
  \bibinfo{author}{\bibfnamefont{P.}~\bibnamefont{Calabrese}},
  \bibnamefont{and} \bibinfo{author}{\bibfnamefont{R.}~\bibnamefont{Fazio}},
  \bibinfo{journal}{J. Stat. Mech.} p. \bibinfo{pages}{0602P001}
  (\bibinfo{year}{2006}).

\end{thebibliography}

\end{document}